\documentclass[preprint,12pt,3p]{elsarticle}

\usepackage[british]{babel}
\usepackage[T1]{fontenc}
\usepackage[utf8]{inputenc}

\usepackage[italic]{mathastext}

\usepackage[sfdefault,condensed]{roboto}
\usepackage{setspace}
\usepackage{soul}

\usepackage[dvipsnames]{xcolor}

\usepackage{geometry}
\usepackage[left,mathlines]{lineno}

\usepackage{graphicx}
\usepackage{booktabs}
\usepackage{tabularx}
\usepackage{adjustbox}
\usepackage{siunitx}
\usepackage[version=4]{mhchem}

\PassOptionsToPackage{hyphens}{url}
\usepackage{hyperref}
\hypersetup{
	colorlinks = true,
	urlcolor   = Orange,
	linkcolor  = RoyalBlue,
	citecolor  = RoyalBlue,
	pdftitle   = {CuCrZr irradiation performance},
	pdfauthor  = {Thomas Barzic and Matheus A. Tunes et al.}
}
\usepackage{xurl}
\biboptions{sort&compress}

\newcommand{\revision}[1]{\textcolor{black}{#1}}
\DeclareSIUnit{\atompercent}{at.\%}
\DeclareSIUnit{\ions}{ions}

\begin{document}
	%\linenumbers
	\begin{frontmatter}
		
		\journal{Acta Materialia}
		
		\title{\texorpdfstring%
			{\raggedright\LARGE\textbf{CuCrZr heat-sink irradiation performance reveals new challenges for thermonuclear fusion reactors}}%
			{CuCrZr heat-sink irradiation performance reveals new challenges for thermonuclear fusion reactors}}
		
		% --- Author list ---
		\author[EEIGM,MUL]{Thomas BARZIC\corref{cor}}
		\ead{thomas.barzic90@gmail.com}
		\author[MUL]{Anna-Carina SEITLINGER}
		\author[MUL]{Christoph FRÜHWIRTH}
		\author[UoB]{Edward MCDONALD}
		\author[MULM]{Jing TANG}
		\author[UKAEA]{Jonathan A. HINKS}
		\author[SCK]{Alexandr ZINOVEV}
		\author[SCK]{Dmitry TERENTYEV}
		\author[MUL]{Stefan LUIDOLD}
		\author[USP]{Cláudio G. SCHÖN}
		\author[UB]{Enrique JIMENEZ-MELERO}
		\author[MUL]{Stefan POGATSCHER}
		\author[MUL,UKAEA]{Matheus A. TUNES\corref{cor}}
		\ead{matheus.tunes@unileoben.ac.at}
		
		\cortext[cor]{Corresponding authors:}
		
		% --- Affiliations ---
		\address[EEIGM]{L'école Européenne d'Ingénieurs en Génie des Matériaux, Université Lorraine, 6 rue Bastien-Lepage F-54010 Nancy, France}
		
		\address[MUL]{Department Metallurgy, Chair of Non-Ferrous Metallurgy, Laboratory of Metallurgy in Extreme Environments, Montanuniversität Leoben, Franz-Josef-Strasse 18, 8700 Leoben, Austria}
		
		\address[UB]{School of Metallurgy and Materials, University of Birmingham, Edgbaston, Birmingham, B15 2TT, United Kingdom}
		
		\address[MULM]{Department Materials Science, Chair of Physical Metallurgy, Montanuniversität Leoben, Franz-Josef-Strasse 18, 8700 Leoben, Austria}
		
		\address[UKAEA]{UKAEA (United Kingdom Atomic Energy Authority), Culham Campus, Abingdon,
			Oxfordshire, OX14 3DB, United Kingdom}
		
		\address[SCK]{SCK CEN, Boeretang 200, 2400 Mol, Belgium}
		
		\address[USP]{Universidade de São Paulo Escola Politécnica, Department of Materials Science and Metallurgical Engineering, Av. Prof. Mello Moraes 2463, 05508-030 São Paulo, Brazil}
		
		\begin{abstract}
			\onehalfspacing
			\noindent Commercial fusion energy requires materials that survive intense neutron bombardment whilst extracting extreme heat loads for conversion to electricity. The CuCrZr alloy, the leading heat-sink material for fusion reactors, derives its strength from a fine dispersion of nano-precipitates formed during prime-ageing heat-treatment. Whether this precipitation-hardening strategy can withstand fusion-relevant irradiation remains untested. \revision{Here we show, combining \textit{in situ} transmission electron microscopy under heavy-ion irradiation and He implantation with thermodynamic and transmutation modelling, that the hardening precipitates dissolve under two opposing kinetic regimes: ballistic dissolution dominates at low temperatures, whilst dissolution and re-precipitation dominate at high temperatures. Although the accelerated dose rates inherent to ion irradiation shift the balance between ballistic mixing and thermal back-diffusion relative to reactor conditions, precipitate degradation at both kinetic extremes indicates that the prime-aged microstructure is unlikely to remain unaltered under prolonged neutron exposure.} He bubbles and Kr-rich voids nucleate once vacancies become mobile, and transmutation over five service years irreversibly redirects the alloy chemistry towards Ni-Zr intermetallics. These three independent mechanisms converge to challenge the strategy on which CuCrZr performance depends, suggesting that the long-term performance of age-hardenable Cu-based heat-sink alloys in fusion reactors warrants further assessment. Our findings reveal a new materials challenge for fusion reactor design and commercialisation: the need for new Cu-based heat-sink alloys able to retain engineered strength whilst their chemistry is irreversibly rewritten -- thermodynamically and ballistically -- by the fusion neutron spectrum.
		\end{abstract}
		
		\begin{keyword}
			Thermonuclear fusion \sep Copper alloys \sep Heat-sinks \sep Radiation damage \sep CuCrZr
		\end{keyword}
		
	\end{frontmatter}

	\newpage
	\tableofcontents
	\newpage
	
	\section{Introduction} % INTRO IS FINISHED OKOK
	\label{sec:introduction}
	\onehalfspacing
	\noindent Understanding and selecting alloys and materials for the development of thermonuclear fusion remains a major challenge for metallurgy given the operating constraints imposed by the multiple degradation forces present in this form of energy technology. One of the major global collaboration projects of our times is the International Thermonuclear Experimental Reactor (ITER) in France. A plasma will reach hundreds of millions of degrees to allow the nuclear fusion reaction between deuterium and tritium atoms in this reactor, thus subjecting the components of the reactor to very high heat fluxes, rapid temperature gradients and an intense stream of energetic particles causing radiation damage. In a fusion reactor, the kinetic energy of fast neutrons ($\approx$ 14 MeV) as well as the electromagnetic radiation from the plasma will serve as sources of energy to be converted into thermal energy. Research on materials for thermonuclear fusion has been intense, yet largely focused on plasma-facing armour materials such as W \cite{el2014situ,el2014ultrafine,yi2015characterisation,yi2016situ,yi2017study}, W-based alloys \cite{yi2013situ,yi2018high,yi2019high,el2020situ} and, more recently, refractory high-entropy alloys \cite{el2019outstanding,el2021helium,el2023quinary}. Comparatively less attention has been devoted to the heat-sink materials immediately behind the armour, despite their pivotal role: they must extract the intense heat loads into the coolant, thereby enabling the conversion of fusion energy into usable electricity.
	
	In terms of materials selection criteria, heat-sink alloys for thermonuclear fusion reactors must exhibit a triad of desirable physical properties: (i) high thermal conductivity, (ii) high strength and (iii) high irradiation resistance -- all at low, moderate and high temperatures. Up to date, the leading candidate to fulfil the heat-sink components of experimental fusion reactors -- such as ITER -- is a terminal solid-solution age-hardenable alloy with Cu featuring Cr and Zr as minor alloying elements \cite{bochvar2007cr,edwards2007effect,engel2025modelling,fabritsiev2005effect,fabritsiev2007effect,gong2024electropolishing_CuCrZr,huang2024corrosion,kalinin2007ageing,kwok2009effect,obitz2016erosion,park2011investigation}. Previously reported criteria defined what are the limits that heat-sink alloys must endure within fusion reactors. In the case of ITER inner components, due to the intense energetic irradiation fluxes, Hirai \textit{et al.} determined that the CuCrZr must withstand a minimum dose of 0.4 displacements-per-atom (dpa) within a narrow operating temperature window from \SI{200}{\degreeCelsius} to \SI{350}{\degreeCelsius} \cite{hirai2016use}, however, overheating conditions to temperatures around of \SI{600}{\degreeCelsius} may be experienced during operation \cite{kalinin2007ageing}, where the alloy must withstanding its designed initial properties. After ITER, heat-sink alloys in future commercial fusion reactors are expected to endure doses up to 20 dpa and higher \cite{EUROfusion2018Roadmap,UKFusionMaterialsRoadmap2040}. 
	
	\subsection{The performance of the CuCrZr in thermonuclear fusion: \textit{Quo imus?}}
	\label{sec:intro:quovadis}
	
	\noindent The CuCrZr is an age-hardenable alloy in which high-strength is attained via precipitation-strengthening \cite{bochvar2007cr,edwards2007effect,engel2025modelling,fabritsiev2005effect,fabritsiev2007effect,gong2024electropolishing_CuCrZr,huang2024corrosion,kalinin2007ageing,kwok2009effect,obitz2016erosion,park2011investigation}. Despite alloying, its thermal conductivity is approximately \SI{320}{\watt\per\metre\per\kelvin} \cite{fabritsiev2005effect} -- compared to \SI{400}{\watt\per\metre\per\kelvin} for pure Cu. In addition, the alloy holds its strength at both at low and at the targeted operational temperatures: Fabritsiev \textit{et al.} reported at both \SI{150}{\degreeCelsius} and \SI{300}{\degreeCelsius}, yield strengths of \SI{260}{\mega\pascal} and \SI{240}{\mega\pascal}, respectively, in unirradiated conditions. More importantly, the CuCrZr alloy is also reported to present some compatibility with W, Be, and selected steels \cite{tejado2018evolution,terentyev2023effect,chen2025effect}, thus allowing cladding via welding to build the composite structure of the plasma walls of ITER \cite{fabritsiev2005effect,terentyev2023effect}. Previous works addressed on the CuCrZr performance in the simulated extreme conditions of thermonuclear fusion reactors. We report here a case-study of the CuCrZr alloy subjected to fusion reactor conditions.
	
	\begin{figure}[ht!]
		\centering
		\includegraphics[width=0.8\linewidth]{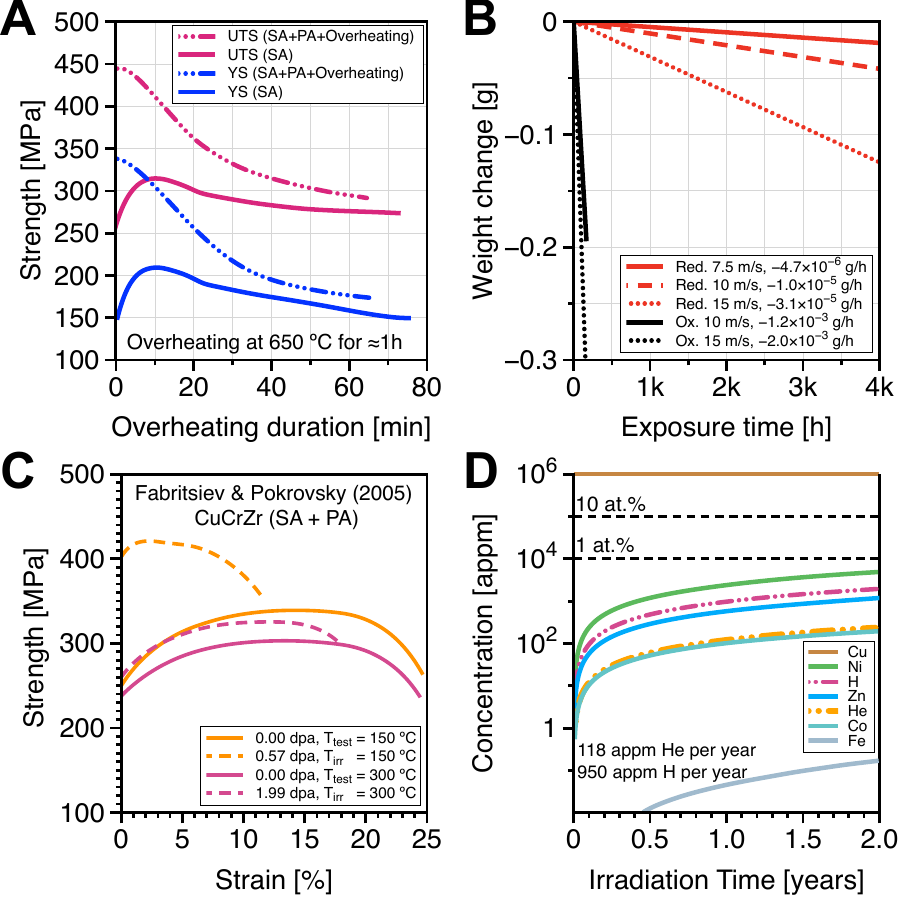}
		\caption{Mechanical properties, corrosion performance, and transmutation yield of the CuCrZr heat-sink alloy. The data in \textbf{A} was replotted from Kalinin \textit{et al.} \cite{kalinin2007ageing} with permission granted by Elsevier: \revision{here overheating was intentionally applied to simulate the effect of ITER abnormal conditions on the heat-sink material}; The data in \textbf{B} was replotted from Obitz \textit{et al.} \cite{obitz2016erosion} with permission granted by Elsevier; The data in \textbf{C} was replotted from Fabritsiev \& Pokrovsky \cite{fabritsiev2005effect} with permission granted by Elsevier; and the data in \textbf{D} was replotted from Gilbert \textit{et al.} \cite{gilbert2015handbook} with permission granted from UK Atomic Energy Authority (UKAEA).}
		\label{fig:LitRev}
		
	\end{figure}
	
	As shown in Fig. \ref{fig:LitRev}A, Kalinin \textit{et al.} investigated the evolution of the mechanical properties of the CuCrZr alloy after primary ageing (PA) and solution annealing (SA) \cite{kalinin2007ageing} in normal and overheating conditions. The latter experiment was conducted at \SI{600}{\degreeCelsius} for 1 h to simulate overheating under ITER-relevant conditions. The total tensile elongation, not included in the figure, is approximately 30\% \cite{hirai2017iter}. It can be noted in Figure \ref{fig:LitRev} that under SA conditions, both the ultimate tensile strength (UTS) and yield strength (YS) are around \SI{200}{\mega\pascal} lower than under PA conditions. Specifically, UTS and YS both decrease from approximately \SI{450}{\mega\pascal} and \SI{350}{\mega\pascal} in the PA state to \SI{250}{\mega\pascal} and \SI{150}{\mega\pascal} in the SA state. Moreover, after (only) one hour of overheating, the YS and UTS of the PA-treated alloy drop to nearly the same level as those of the SA condition. Given the drastic reduction of both YS and UTS, it is evident that under overheating conditions, CuCrZr undergoes a severe loss of mechanical strength due to its limited thermal stability. 
	
	Obitz \textit{et al.} employed a submerged jet impingement technique to simulate the primary cooling circuit of the ITER reactor \cite{obitz2016erosion}, enabling assessment of the corrosion behaviour of CuCrZr. As shown in Fig. \ref{fig:LitRev}B, the mass-loss rate at \SI{200}{\degreeCelsius} is substantial in an oxidising environment (the case in ITER). Independent work by Huang \textit{et al.} confirmed pronounced erosion--corrosion in CuCrZr, reporting high degradation rates under similar conditions \cite{huang2024corrosion}. Compared with other nuclear materials tested in the same regime, this alloy exhibits markedly higher corrosion rates. During tests lasting (only) 166 days, several degradation mechanisms were observed. Variations in electrochemical potential -- typical during plasma operation -- led to severe corrosion at elevated temperatures. Although stable reducing conditions mitigated the rate, damage continued to accumulate with increasing exposure time. Despite the relatively short duration, these findings demonstrate that CuCrZr is highly susceptible to corrosion in fusion-relevant environments. To date and to be best of our knowledge, no experiments have examined the combined effects of irradiation and corrosion simultaneously for this specific alloy; however, ionising radiation and energetic particles are well known to accelerate corrosion in nuclear materials \cite{bjorkbacka2013radiation,raiman2017accelerated,liu2024review}.
	
	Irradiation campaigns were carried out in materials test reactors both in Russia (SM‑2 and SPP‑6/7 facilities) and Belgium (BR2). Fabritsiev and Pokrovsky \cite{fabritsiev2005effect} as well as Edwards \textit{et al.} \cite{edwards2007effect}, performed most of the fast neutron irradiation studies on the CuCrZr alloy found in the literature. Fabritsiev and Pokrovsky \cite{fabritsiev2005effect} reported for the PA condition, a strong increase in yield strength together with a marked reduction in ductility already at relatively low doses and irradiation temperatures. These experiments were performed up to 0.54 dpa under neutron irradiation at \SI{150}{\degreeCelsius}. At this dose, the uniform elongation decreased from 25\% in the unirradiated condition to 10\% when tested at the same temperature. Under irradiation at temperatures around \SI{300}{\degreeCelsius} and up to 1.99 dpa, the plasticity of the alloy is reported to partially recover. It is worth noting that such large differences in mechanical behaviour arise from a temperature change of only \SI{150}{\degreeCelsius} (T$_{m}$=0.3). Furthermore, Edwards \textit{et al.} showed that in an over-aged condition, CuCrZr exhibits better work‑hardening capacity and, at doses of around 0.3 dpa, less strain localisation \cite{edwards2007effect}. Although over-ageing only partially mitigates irradiation damage, CuCrZr in the over-aged condition retains better mechanical behaviour after irradiation than in the PA condition, although the achievable strain levels remain low.
	
	The effect of neutron irradiation on thermal conductivity was also investigated by Fabritsiev and Pokrovsky \cite{fabritsiev2005effect}. Up to 2 dpa, the thermal conductivity decreases to about \SI{200}{\watt\per\metre\per\kelvin} and \SI{225}{\watt\per\metre\per\kelvin} for irradiation at \SI{150}{\degreeCelsius} and \SI{300}{\degreeCelsius}, respectively, compared with approximately \SI{320}{\watt\per\metre\per\kelvin} in the initial SA + PA state. Such a reduction of around 40\%, combined with irradiation damage and the high operating temperatures in fusion, clearly degrades the alloy's ability to remove heat from the fusion reactor.
	According to calculations performed by Gilbert \textit{et al.} \cite{gilbert2015handbook} and presented herein in Fig. \ref{fig:LitRev}C, the CuCrZr alloy activates under neutron irradiation, leading to transmutation into metals such as Ni, Zn, Co and Fe. In long-term, transmutation will eventually contribute to degradation of both mechanical and thermal properties, but this hypothesis is yet to be tested. In addition, as calculated by Gilbert \textit{et al.} \cite{gilbert2015handbook}, Cu transmutation can accumulate around 950 appm of H and 200 appm of He within one year of service in a fusion reactor. This accumulation can cause severe embrittlement via the nucleation of nano-cavities such as H$_2$+He bubbles that will ultimately accelerating cracking that can compromise the structural integrity of the heat-sink pipelines \cite{tunes2024limitations}. It is worth emphasising that the gaps in knowledge on the effects of transmutation elements in fusion materials have been largely overlooked in the past recent years, as recently noted by Reali \textit{et al.} \cite{reali2026teorysimulation}.
	
	\subsection{Objectives of this work}
	\label{sec:intro:obj}
	\noindent The trade-off between high strength, high thermal conductivity and high corrosion resistance in Cu-based alloys -- especially the ternary CuCrZr variants \cite{bochvar2007cr,edwards2007effect,engel2025modelling,fabritsiev2005effect,fabritsiev2007effect,gong2024electropolishing_CuCrZr,huang2024corrosion,kalinin2007ageing,kwok2009effect,obitz2016erosion,park2011investigation} -- gained momentum in the past five years \cite{zhang2020nano,yang2021nano,xu2018CuCrAg,wang2022underaging,hu2022twostageaging,liang2024nano,zhang2025overcoming} due to the universal aspiration of deploying a commercial fusion reactor amid worldwide climate changes and energy crises. Yet, detailed studies showing how energetic particle irradiation changes the microstructure of the heat-sink CuCrZr alloys remain scarce, and the existing neutron irradiation database was largely acquired in fission reactor spectra, which cannot access the threshold (n,$\alpha$) and (n,p) transmutation reactions characteristic of the fusion neutron spectrum and therefore overlook the transmutation-driven degradation channels examined herein. In this work, we perform a study featuring \textit{in situ} transmission electron microscopy (TEM) with heavy-ion irradiation and \textit{in situ} TEM with He implantation on the CuCrZr alloy in order to fully explore the effects of the expected radiation damage that this heat-sink alloy may experience in future thermonuclear fusion reactors. The heavy ions directly emulate the large and dense defect cascades observed when fast fusion neutrons ($\approx$ \SI{14}{\mega\electronvolt}) impinge upon the materials within the fusion reactors, whilst serving as an accelerated surrogate for neutron damage: the ion dose rate exceeds the calculated in-service displacement rate by several orders of magnitude, thereby enabling dose levels equivalent to years of reactor operation to be reached and monitored in real time. The He implantation simulates the behaviour of this light nucleus within the heat-sink alloy when Cu transmutes into alpha particles under neutron bombardment; the atomic-percent concentrations implanted herein emulate the cumulative transmutant inventory accrued over prolonged service. The study investigates irradiation and He implantation across low- and high-temperature regimes to evaluate the kinetics of radiation damage and the microstructural evolution of defects under immobile and mobile conditions. Room-temperature experiments provide the immobile-defect baseline, whilst experiments at \SI{650}{\degreeCelsius} deliberately exceed the upper operational limit of CuCrZr in ITER (\SI{350}{\degreeCelsius}) and the \SI{450}{\degreeCelsius} envisaged for DEMO, thereby accelerating the thermally activated defect kinetics expected over prolonged service. \revision{The two irradiation temperatures were selected to bracket the fusion-relevant operating window, isolating the athermal and thermally-activated degradation pathways whose superposition is expected in service.} Our study comprises a full characterisation of the CuCrZr alloy before irradiation, during irradiation with \textit{in situ} TEM, and complete post-irradiation characterisation with scanning transmission electron microscopy (STEM) coupled with analytical techniques such as energy dispersive X-ray (EDX) spectroscopy. A discussion on the equilibrium of the CuCrZr before and after irradiation is carried out considering both FactSage and FISPACT-II calculations.
	
	\section{Materials and methods} % MATMET IS FINISHED OKOK
	\label{sec:matmet}
	
	\subsection{Chemical analysis}
	\label{sec:metmat:chemanalysis}
	\noindent The CuCrZr alloy used in this work was received from the SCK CEN laboratory in Belgium and was manufactured by Thyssen Duro Metal GmbH in Germany. The manufacturer provided chemical analysis of the alloy as follows (in wt.\%): 0.67Cr, 0.083Zr with Cu balance (Cu--99.11 at.\%
	Cr--0.82 at., Zr--0.06 at.\%). We received the batch number 4 of samples for which the data-sheet from the manufacturer can be found in the supplemental information (SupInfo1). The alloy was received in the prime-aged state after hot rolling, denoted herein as PA.
	
	\subsection{Sample preparation}
	\label{sec:metmat:sampleprep}
	\noindent The CuCrZr alloy was first sectioned using the Struers cutting machine Secotom 15. Then they were cold-embedded in \SI{30}{\gram} of epoxy resin. The trapped air was removed with a vacuum impregnation system for \SI{7}{\minute} and the mountings were kept under pressure for 2 days to ensure proper curing. After mounting, the samples were ground and polished using the Struers Tegramin 30 machine. Oxide-Polish Suspension (OPS) suspension (colloidal silica) was applied at the end of the polishing sequence in order to obtain a suitable scratch-free mirror surface for electron microscopy.
	
	\subsection{Heat-treatments}
	\label{sec:metmat:heat-treatments}
	\noindent Sectioned samples were subjected to a solution heat-treatment at \SI{980}{\degreeCelsius} for 30~minutes in a nitrogen gas (N$_2$) atmosphere followed by water quenching. Then they were again PA at \SI{475}{\degreeCelsius} for a duration between 1 and 4~h, after which they were left to cool in air. \cite{hirai2017iter,kalinin2007ageing}.
	
	\subsection{Hardness measurements}
	\label{sec:metmat:Hardness}
	\noindent The hardness of each sample was measured in three key stages: (i) in pristine condition (as-received from SCK), (ii) after solution annealing and water quenching, and (iii) after subsequent heat-treatment until the PA condition. The hardness was determined using the Brinell hardness tester Emcotest M4C 025 G3M with an indenter ball of \SI{2.5}{\milli\metre}, with a testing time of \SI{15}{\second} and a force of \SI{613.13}{\newton}. For statistics, three indentations were performed for each condition to obtain an accurate average value.
	
	\subsection{Scanning electron microscopy}
	\label{sec:metmat:SEM}
	\noindent Scanning electron microscopy (SEM) was performed using a JEOL JSM IT300LV microscope. The specimens were prepared as described in the \ref{sec:metmat:heat-treatments} and a copper strip was placed on each sample to improve electrical grounding. Secondary electron and backscattered electron images were acquired. Energy Dispersive X-ray (EDX) spectroscopy analyses were carried out using an Oxford Instruments Ultim Max 100 detector.
	
	\subsection{Phase quantification and identification with ESPM}
	\label{sec:metmat:ESPM}
	
	\noindent Phase identification and compositional quantification of precipitates were performed directly on the STEM-EDX data generated in this research by using a custom-implemented graphical user interface based on the open-source Python package ESPM (electron spectro-microscopy) developed by Teurtrie \textit{et al.} \cite{teurtrie2023espm}. The package decomposes STEM-EDX hyperspectral datacubes via non-negative matrix factorisation (NMF), yielding spectral components that correspond to distinct phases alongside their spatial abundance maps. Elemental concentrations in atomic percent are derived by converting fitted spectral weights using X-ray emission cross-sections from the support of the pre-existing \texttt{emtables} package. This workflow enabled simultaneous phase segmentation and local chemical quantification of precipitates across the acquired datasets. Both packages are freely available at \href{https://github.com/adriente/espm}{github.com/adriente/espm} and \href{https://github.com/adriente/EMTables}{github.com/adriente/EMTables}.
	
	\subsection{Electron-backscattered diffraction}
	\label{sec:metmat:EBSD}
	\noindent Electron backscatter diffraction (EBSD) was conducted using a Tescan Clara field-emission SEM equipped with an Oxford Symmetry S3 detector. It is worth emphasising that the sample preparation method described in \ref{sec:metmat:sampleprep} was sufficient to obtain appropriate EBSD patterning. The utilised acceleration voltage and probe current were \SI{20}{kV} and \SI{10}{nA}, respectively. The step size for data collection was \SI{200}{nm}. Data analysis was performed within the software Aztec Crystal directly on the raw data.
	
	\subsection{Twin jet electropolishing}
	\label{sec:metmat:TJE}
	\noindent After sample preparation as described in subsection \ref{sec:metmat:sampleprep}, \SI{3}{\milli\metre} disks were punched out of the samples and their thicknesses were around \SI{80}{\micro\metre}. A Struers TenuPol-5 was used for Twin Jet Electropolishing (TJE) in order to obtain electron-transparent samples for \textit{in situ} ion irradiation within TEM. For that, an electrolyte composed of 33\% nitric acid and 66\% methanol (in vol.\%) was used at an anodic potential of \SI{5}{\volt} and the temperature of the bath was held constant at \SI{-40}{\degreeCelsius}. After TJE, the samples were mirror-like with a hole in the centre. A similar TJE recipe was reported by Gong \textit{et al.} \cite{gong2024electropolishing_CuCrZr}.
	
	\subsection{Scanning transmission electron microscopy}
	\label{sec:metmat:STEM}
	\noindent The pre- and post-irradiated samples were analysed in a Thermo Fisher Talos STEM operating an X-FEG source at \SI{200}{\kilo\electronvolt}. This microscope is located at the Montanuniversität Leoben in Austria and it is equipped with all standard STEM detectors, comprising Bright-Field (BF-STEM), Dark-Field (DF-STEM), and High-Angle Annular Dark-Field (HAADF) as well as a Ceta Camera (CMOS technology) for conventional bright-field TEM (BFTEM) and through-focal imaging. Selected-Area Electron Diffraction (SAED) and High-Resolution TEM (HRTEM) were also used to characterise the nanometre-sized precipitates structures and to track the alloy evolution under irradiation. The microscope also features the Super-X EDX detectors for fast throughput analytical data acquisition and mapping, which was used for pre- and post-irradiation characterisation. \revision{For the EDX mapping experiments, the acceleration voltage was preserved at \SI{200}{\kilo\electronvolt} and the probe current was around of \SI{1}{\nano\ampere}.}
	
	\begin{figure}[ht!]
		\centering
		\includegraphics[width=0.7\linewidth]{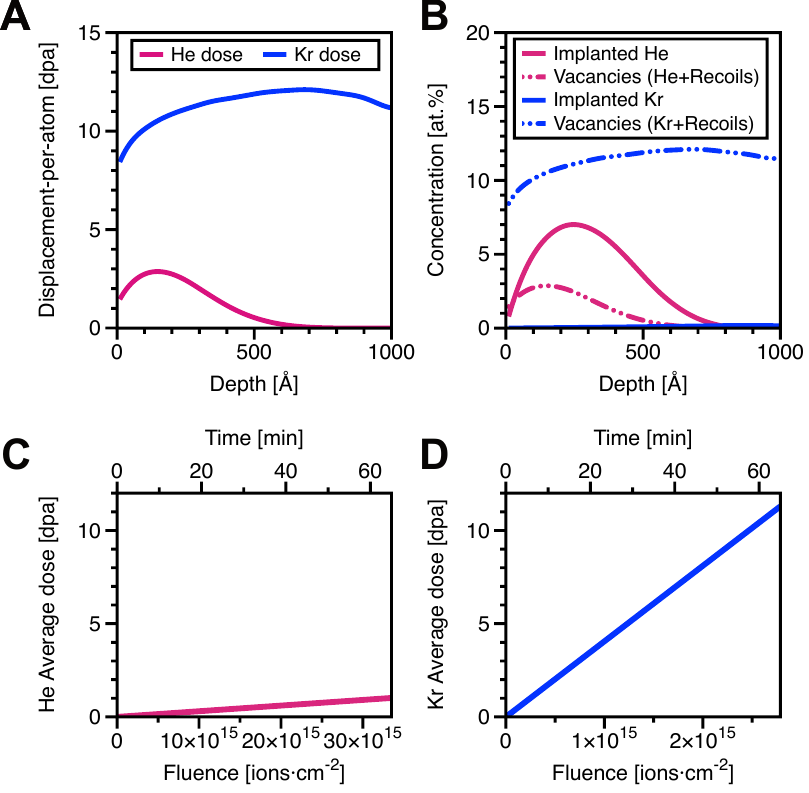}
		\caption{Radiation damage calculations for both heavy-ion (Kr, blue) irradiation and light-ion (He, magenta) implantation on the CuCrZr heat-sink alloy performed with SRIM-2013Pro \cite{ziegler2010srim}. \textbf{A} shows the depth profiles of the displacement damage, \textbf{B} the depth-dependent atomic concentration profiles (solid line) and the total vacancies' (dashed line) distributions at the maximum fluences: for He \SI{3.4e16}{ions\per\centi\metre\squared} and for Kr \SI{2.8e15}{ions\per\centi\metre\squared}. \textbf{C} and \textbf{D} display the evolution of the average damage dose (over \SI{100}{\nano\metre}) as a function of both irradiation time and fluence for He and Kr irradiations, respectively. Note: For better visualisation, the Monte Carlo simulations from SRIM-2013Pro used 10000 ions and the curves were smoothed using a LOESS (locally estimated scatterplot smoothing) algorithm.}
		\label{fig:IrradiationProfiles}  
	\end{figure}
	
	\subsection{\textit{In situ} ion irradiation and implantation at the UKAEA's MIAMI-2 Facility}
	\label{sec:metmat:MIAMI}
	
	\noindent The \textit{in situ} TEM heavy-ion irradiation and implantation have been carried out at the MIAMI-2 Facility operated by the United Kingdom Atomic Energy Authority \cite{greaves2019new}. For the heavy-ion irradiations, a \SI{600}{\kilo\electronvolt} Kr$^{+2}$ ion beam was used whereas for the He implantations, a \SI{6}{\kilo\electronvolt} He$^{+}$ ion beam was employed. The physical basis for employing \SI{600}{\kilo\electronvolt} Kr$^{+2}$ ions to emulate neutron irradiation damage in the CuCrZr alloy is grounded in the PKA energy spectrum calculated by Gilbert \textit{et al.}~\cite{gilbert2015handbook} for Cu under a fusion neutron flux. The spectrum reveals that Cu PKAs span a broad energy range, with the rate distribution peaking at energies approaching \SI{\sim 1}{\mega\electronvolt}. At \SI{600}{\kilo\electronvolt}, the incident Kr$^{+2}$ ions transfer recoil energies to Cu lattice atoms commensurate with the high-energy tail of this PKA spectrum, thereby generating dense displacement cascades of comparable character to those produced by \SI{14}{\mega\electronvolt} fusion neutrons -- precisely the damage morphology that heavy ions are well suited to replicate. It is worth commenting that G. Was positively discussed the use of ion beams to emulate neutron damage in thermal nuclear reactors \cite{was2015challenges} and our research group has been recently applying such a methodology aforementioned specifically for fusion materials \cite{tunes2019d_transmission,tunes2020transmission,tunes2022accelerated}.
	
	In MIAMI-2, the damaged microstructures were live-monitored as a function of irradiation/implantation time using a Hitachi H-9500 TEM operating a LaB$_6$ filament at \SI{300}{\kilo\volt}. This TEM has a point-to-point spatial resolution of 0.18 nm. Both ion beams impinged at the specimen position at an angle of \SI{18.5} relative to the electron beam. The heavy-ion flux was \SI{7.7e11}{\per\centi\metre\squared\per\second} whereas the He flux was \SI{9.3e12}{\per\centi\metre\squared\per\second}. The final fluence for each cases were: for He \SI{3.4e16}{ions\per\centi\metre\squared} and for Kr \SI{2.8e15}{ions\per\centi\metre\squared}. It is worth emphasising that the ion beam fluxes were measured at the specimen position using a custom-implemented Current Meter Rod (CMR) as described in the MIAMI-2's technical publication \cite{greaves2019new}. The high-temperature irradiations and implantations have been carried out using a Gatan furnace heating holder Model 652 double-tilt. Damage calculations as well as fluence-to-dpa conversion is shown in Fig. \ref{fig:IrradiationProfiles}A-D: for that, we have used a procedure suggested by Stoller \textit{et al.} \cite{stoller2013use} which uses the Kinchin-Pease quick-mode of calculation. For these calculations, the alloy density was assumed to be \SI{8.9}{\gram\per\cubic\centi\metre} and a range of \SI{1000}{\angstrom} was set in SRIM. Displacement damage energies were default 25 eV for Cu, Cr, and Zr. He concentrations in this paper were estimated considering the average implanted ions over a \SI{100}{\nano\meter} foil. \revision{Following the Stoller \textit{et al.} procedure Both lattice and surface binding energies were set to zero.}
	
	\subsection{Equilibrium thermodynamics}
	\label{sec:metmat:FactSage}
	\noindent Thermodynamic equilibrium calculations were performed with the FactSage 8.4 software package applying the SGTE 2020 database \cite{bale2002factsage,bale2009factsage}. These calculations serve as a basis to interpret the (equilibrium) microstructure of the as-received CuCrZr heat-sink alloy both at meso- and nano-scales as-received, before and after heat-treatment, and before and after irradiation. 
	
	\subsection{Transmutation calculations with FISPACT-II}
	\label{sec:metmat:FISPACT}
	\noindent Transmutation calculations using FISPACT-II v5.1 were performed assuming the nominal composition of the CuCrZr alloy introduced in subsection \ref{sec:metmat:chemanalysis}, this was included in the input file as weight percentages \cite{sublet2017fispact}. These computations employed two nuclear data libraries. TENDL-2017 was used for its cross-section data and the unresolved resonance region (URR) probability tables, whilst the \textsc{Decay2020} dataset was utilised for the nuclear index file and its decay data \cite{nuclib}. URR data was required to account for energy self-shielding arising from unresolved resonances in the cross-section data, allowing for more accurate reaction rates to be calculated \cite{ProbTab}. A representative neutron flux spectra for a D-T fusion reactor with a He-cooled pebble bed (HCPB-FW), provided by the UKAEA, was used for these transmutation calculations \cite{FISPACTwiki}. This spectra provided a flux of 6.60 $\times$ 10$^{14}$ neutrons cm$^{-2}$ s$^{-1}$.
	
	It is worth emphasising that these transmutation calculations were performed for 1 kg of the CuCrZr alloy for a time frame spanning five full power years. This resulted in a fluence of 1.04 $\times$ 10$^{23}$ neutrons cm$^{-2}$, which is $\approx$ 50 dpa: not realistic for ITER, but relevant for DEMO and beyond \cite{EUROfusion2018Roadmap,UKFusionMaterialsRoadmap2040}. These calculations supplied the composition of the material at several points throughout this time frame. These values were then used to construct an animation to show the evolution of isotopes present within the material and to discern its final elemental composition. In these animations (see supplemental information: files SupInfo2, 3, and 4) the initial isotopes of Cu, Cr and Zr were marked with a star to aid observation.
	
	\section{Results and discussion}
	\label{sec:resdis}
	
	\subsection{Alloy characterisation in prime-aged as-received condition}
	\label{sec:resdis:alloyPAasreceived}
\begin{figure}[ht!]
		\centering
		\includegraphics[width=\linewidth]{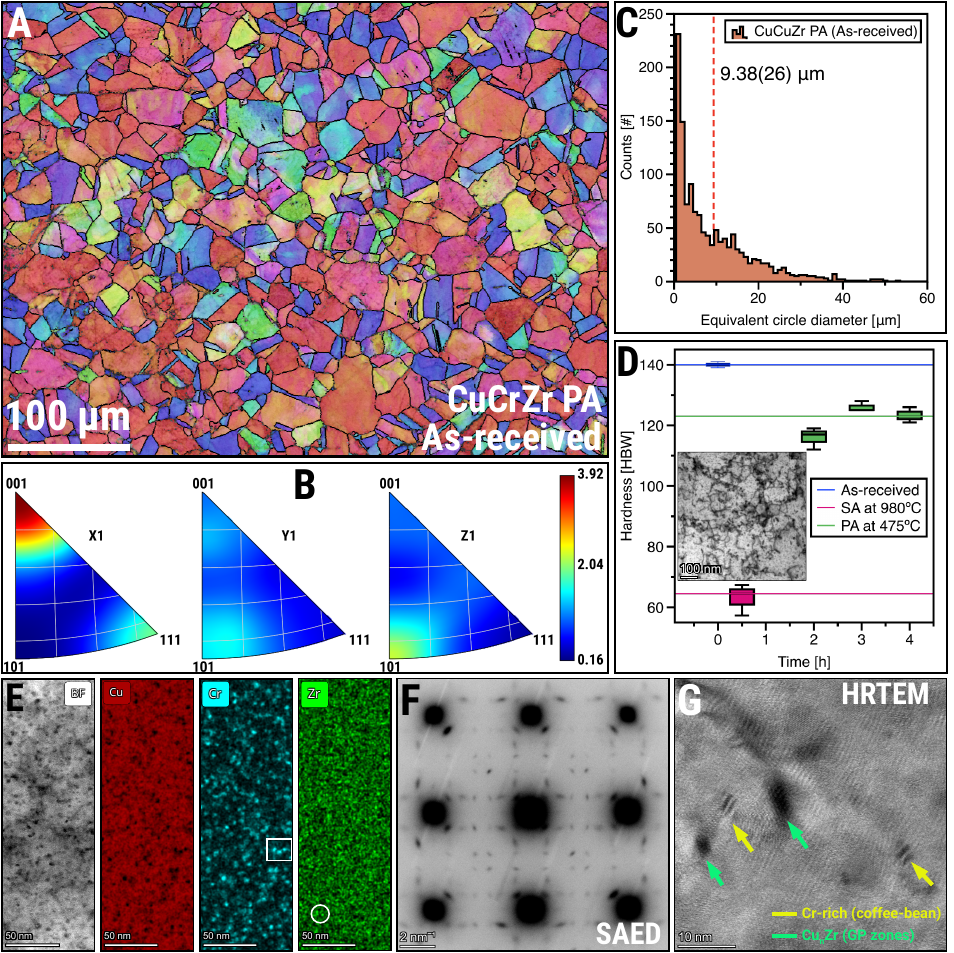}
		\caption{Pristine analysis of the CuCrZr PA alloy as received from SCK. \textbf{A} Shows the IPF Colouring micrograph along the parallel $x$ direction, where the \textbf{B} inverse pole figures show a texture component along the same direction, and a minor texture along $z$ (perpendicular to the sample's surface). \textbf{C} The average grain-size of the alloy was estimated to be \SI{9.38 \pm 0.26}{\micro\metre}. The plot in \textbf{D} shows the evolution of hardness from the as-received, SA at \SI{980}{\degreeCelsius}, and PA at \SI{475}{\degreeCelsius}: the original PA state could not be restored as the as-received alloy has a degree of plastic deformation, consistent with the EBSD data. The STEM-EDX maps in \textbf{E} show the alloy microstructure in the PA as-received condition with nanometre-sized precipitates of two variants, Cr-rich (coffee-bean) and Zr-rich (GP zones). Complimentary SAED \textbf{F} and HRTEM \textbf{G} further characterise the age-hardenable structure.}
		\label{fig:Pristine}
	\end{figure}

	\noindent The as-received CuCrZr PA alloy was fully characterised prior to the irradiation study with a combination of advanced multiscale techniques, such as EBSD, hardness, and STEM-EDX as shown in Fig. \ref{fig:Pristine}.  
	
	%Fig. Pristine CuCrZr PA as-received.
	
	The EBSD assessment shown in the inverse pole figure (IPF) coloured micrograph in Fig. \ref{fig:Pristine}A (\textit{i.e.} IPF-X) reveals an equiaxed-grained microstructure with a minor texture component along the $z$ axis as quantified in Fig. \ref{fig:Pristine}B. In a PA condition, the CuCrZr alloy exhibits the combination of equiaxed grains and a moderate $\{001\}$ cube texture ($\approx$3.92 MRD) alongside a low twin-density. While pure Cu typically develops abundant annealing twins ($\Sigma3$ boundaries), their scarcity here is attributed to retarded grain-boundary migration caused by solute drag from dissolved Cr and Zr~\cite{cahn1962} and pinning by early-stage (Cr,Zr)-rich nano-precipitates~\cite{chen2018}. Since annealing twins are generated by growth at migrating boundaries, their density scales with boundary velocity and migration driving force \cite{pande1990,bozzolo2020}: any impediment to boundary motion during solution annealing therefore suppresses twin formation. The EBSD data also allow quantification of the grain microstructure of the CuCrZr PA alloy. The as-received material exhibits a refined, equiaxed grain structure with an average grain size of \SI{9.38(26)}{\micro\meter} (Fig.~\ref{fig:Pristine}C) and the moderate $\{001\}$ cube texture of $3.92$ MRD suggests some degree of plastic deformation inherited from hot rolling. Such a fine-grained (also suppressing twinning), equiaxed microstructure is desirable for fusion heat-sink applications, combining high thermal conductivity with the grain-boundary and precipitation strengthening required of high-heat-flux pipework.
	
	Starting from the PA as-received condition, hardness evolution was examined following SA and a subsequent attempt to reproduce the PA state via the thermal treatment of Kalinin \textit{et al.} \cite{kalinin2007ageing}. The aim was to determine whether alloy strength arises solely from precipitation hardening, or whether prior plastic deformation also contributes -- as suggested by the EBSD assessment in Fig.~\ref{fig:Pristine}A--B. As shown in Fig. \ref{fig:Pristine}D, SA at \SI{980}{\celsius} reduces hardness from 140 to 65~HBW; subsequent ageing at \SI{475}{\celsius} for 2--4~h did not restore the as-received PA hardness. This indicates that elevated hardness arises not from precipitation strengthening alone, but also from prior plastic deformation -- consistent with the cold-rolling strengthening mechanisms reported by Gong \textit{et al.} \cite{gong2024electropolishing_CuCrZr}. This is further corroborated by the BF-STEM micrograph inset in Fig. \ref{fig:Pristine}D, taken from the as-received PA hot-rolled sample, in which a dense dislocation network is visible.
	
	At the nanoscale, the CuCrZr PA alloy features what appears to be two variants of nanoprecipitates as shown by the BFTEM and elemental maps presented in Fig. \ref{fig:Pristine}E: throughout this work, we refer to these as Cr- and Zr-rich nano-precipitates. When taken along the $[001]$ zone axis, the SAED pattern in Fig. \ref{fig:Pristine}F is consistent with the FCC matrix of Cu. In addition, the presence of satellite reflexions is attributed to superlattice spots due to the presence hardening nano-precipitates. It is worth emphasising that, in this work, the SAED pattern shown in Fig. \ref{fig:Pristine}E establishes a baseline for the CuCrZr PA before ion irradiation. To further corroborate the existence of two variants of hardening precipitates, the HRTEM micrograph in Fig. \ref{fig:Pristine}G exhibit two types of contrast for the hardening precipitate: one resembling to rounded GP zones (the Zr-rich variant or Cu$_{n}$Zr) and one with a coffee-bean morphology (the Cr-rich variant). These STEM-EDX, SAED and HRTEM observations point that the PA heat-treatment produces a metastable microstructure (see later discussion in section \ref{sec:overall:equilibrium}). 
	
	\subsection{Heavy-ion irradiation at low and high temperatures}
	\label{sec:resdis:heavyionirradiation}
\begin{figure}[ht!]
		\centering
		\includegraphics[width=\linewidth]{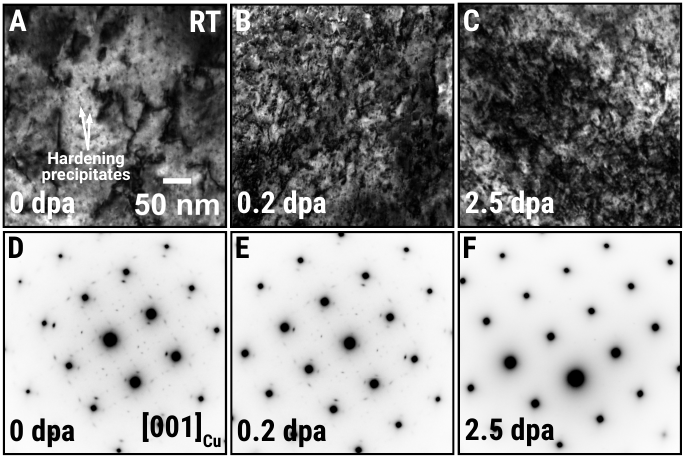}
		\caption{The microstructure of the CuCrZr PA alloy irradiated by 600 keV Kr ions at room temperature showing BFTEM \textbf{A-C} and SAED patterns \textbf{D-F} from 0 to 2.5 dpa. Under irradiation, the alloy accumulates displacement damage within its microstructure in the form of black-spots with concomitant lost in the diffraction signal of the nanometre-sized precipitates. Note: The scale bar in \textbf{A} also applies to \textbf{B} and \textbf{C}.}
		\label{fig:Kr:RT}
	\end{figure}

\begin{figure}[ht!]
		\centering
		\includegraphics[width=0.90\linewidth]{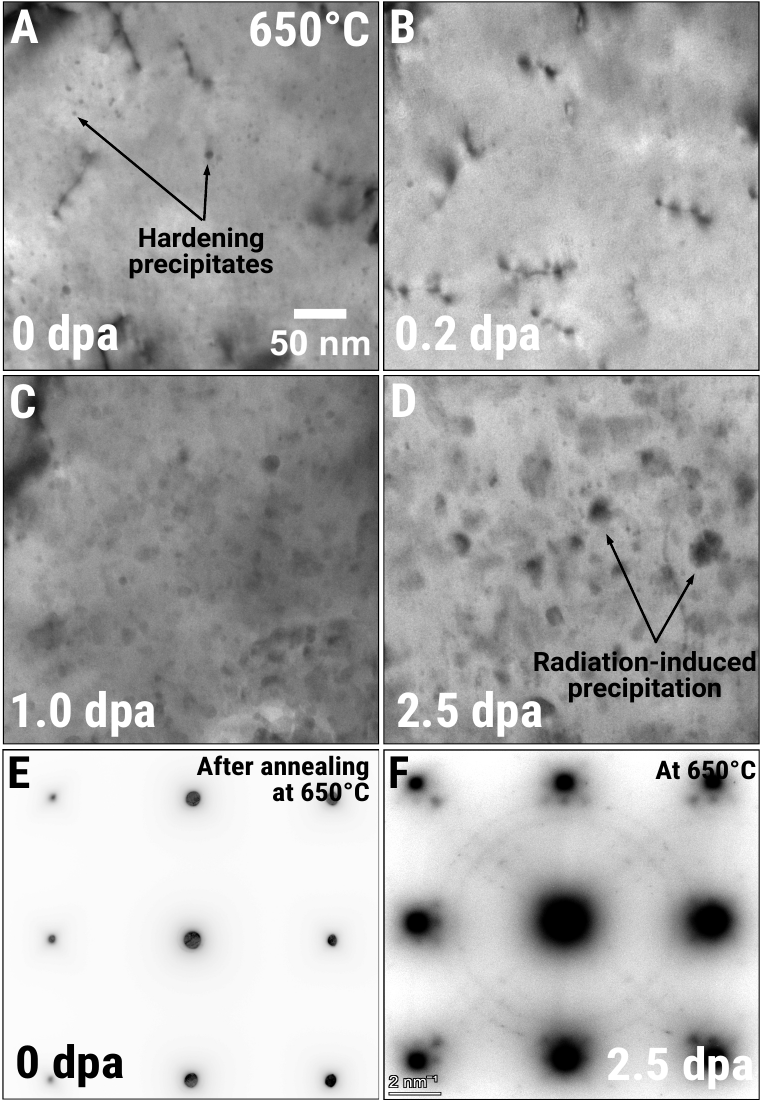}
		\caption{The microstructure of the CuCrZr alloy irradiated by 600 keV Kr ions after heating and hold at \SI{650}{\degreeCelsius} versus damage dose. The images \textbf{A},\textbf{B},\textbf{C} and \textbf{D} illustrate the evolution of the alloy microstructure under BFTEM as a function of damage dose, ranging from 0 to 2.5 dpa. The SAED micrographs \textbf{E} and \textbf{F} highlight changes in the alloy microstructure as-captured via electron diffraction at 0 dpa after annealing without irradiation at \SI{650}{\degreeCelsius} and after irradiation at \SI{650}{\degreeCelsius} up to 2.5 dpa. Note: The scale bar in \textbf{A} also applies to \textbf{B}, \textbf{C}, and \textbf{D}.} 
		\label{fig:Kr:650C}
	\end{figure}

	\noindent The response of the CuCrZr PA alloy to 600 keV Kr$^{+2}$ heavy-ion irradiation was examined with \textit{in situ} TEM under two thermal regimes: room temperature ($\approx$\SI{22}{\degreeCelsius}, $T/T_\mathrm{m} \approx 0.22$, calculated in absolute K) and \SI{650}{\degreeCelsius} ($T/T_\mathrm{m} \approx 0.68$), over a damage range from 0 to 2.5 dpa as shown in Figs.~\ref{fig:Kr:RT} and \ref{fig:Kr:650C}, respectively. The high-temperature was achieved in \SI{10}{\minute} of annealing within \textit{in situ} TEM at MIAMI-2.
	
	%Fig. Kr irrad at RT.
	
	At room temperature, the BFTEM micrographs in Fig.~\ref{fig:Kr:RT}A--C reveal a progressive accumulation of displacement damage in the form of black-spot defects -- small, unresolved defect clusters comprising vacancy and interstitial loops produced by collision cascades -- whose number density fast increases with dose. At this temperature, point defects and small clusters are largely immobile, so damage accumulates in a form of black-spots without (significant) recombination or coarsening. The concomitant changes in the SAED patterns (Fig. \ref{fig:Kr:RT}D--F) are particularly informative: the superlattice reflections associated with the hardening nano-precipitates, clearly visible in the unirradiated baseline (Fig. \ref{fig:Kr:RT}D), progressively weaken and ultimately vanish with increasing dose. This loss of diffraction contrast indicate both radiation-induced partial dissolution of the Cr- and Zr-rich precipitates and/or removal of their long-range crystal symmetry. Ballistic mixing within displacement cascades is a well-established phenomenon \cite{brinkman1954cascades,fry1956cascades,silcox1959cascades,gibson1960cascades,english1976cascades,nordlund1997cascades,nordlund1998cascades,greaves2013sputtering}: upon dissolution, the precipitates' constituents are driven back to the matrix solid-solution. Interestingly, this latter observation -- on energetic particle causing modifications/disruption of nano-hardening precipitates -- has been extensively reported in the context of age-hardenable aluminium alloys within extreme environments \cite{lohmann1987amicrostructure,tunes2020prototypic,willenshofer2026radiation}. 
	
	The response of the CuCrZr PA alloy to high temperatures is markedly different, as shown in Fig. \ref{fig:Kr:650C}. At a temperature of \SI{650}{\degreeCelsius}, the enhanced mobility of both vacancies and interstitials promotes efficient mutual recombination, suppressing the accumulation of black-spot defects observed at room temperature. An important observation is the effect of temperature (without irradiation) on the hardening precipitates. As exhibited by the BFTEM micrograph in Fig. \ref{fig:Kr:650C}A and the SAED pattern in Fig. \ref{fig:Kr:650C}E, both taken directly after annealing at \SI{650}{\degreeCelsius}, the population of hardening precipitates significantly decreased (the remaining precipitates are coarsened) and the diffraction signal attributed to the hardening precipitates in the SAED pattern (superlattice reflexions) vanished, suggesting that these Cr- and Zr-rich nano-phases are not stable at temperatures as high as \SI{650}{\degreeCelsius}. The latter observation is also consistent with the softening effect observed by Kalinin \textit{et al.} \cite{kalinin2007ageing} who performed tensile experiments during overheating conditions typically found in thermonuclear fusion reactor conditions (see replotted data in Fig. \ref{fig:LitRev}A). 
	
	During irradiation at \SI{650}{\degreeCelsius}, the BFTEM sequence in Fig.~\ref{fig:Kr:650C}A--D shows that, rather than a progressive build-up of fine damage clusters (as observed for the low-temperature case in Fig. \ref{fig:Kr:RT}), the microstructure evolves through coarsening and rearrangement of defects into larger, more stable configurations, consistent with thermally activated recovery. A comparison of the SAED patterns before (Fig. \ref{fig:Kr:650C}E) and after 2.5 dpa (Fig. \ref{fig:Kr:650C}F) at \SI{650}{\degreeCelsius} further illustrates the distinct behaviour at elevated temperatures: the electron diffraction signal undergoes changes that reflect both thermally- and radiation-driven microstructural evolution, including precipitate dissolution followed by re-precipitation, \textit{i.e.} radiation-induced precipitation (RIP), which is driven by the interplay between ballistic disordering and thermally- and displacive-activated diffusion at this temperature.
	
	The heavy-ion irradiations at both room temperature and at \SI{650}{\degreeCelsius} indicate that, in summary, the CuCrZr PA alloy is highly susceptible to displacement damage accumulation across both thermal regimes, yet through distinct microstructural pathways. At room temperature, defect mobility is clearly suppressed, leading to a progressive build-up of black-spot damage clusters and radiation-induced dissolution of the hardening nano-precipitates -- a combined degradation of both the matrix and the precipitate microstructure. At \SI{650}{\degreeCelsius}, enhanced defect mobility promotes recovery and coarsening, yet the elevated homologous temperature ($T/T_\mathrm{m} \approx 0.68$) simultaneously drives thermally assisted microstructural evolution that compounds the radiation-induced changes. In both cases, the initial metastable PA microstructure -- optimised for precipitation hardening -- is destabilised by irradiation, with implications for the mechanical performance and thermal conductivity of the heat-sink alloy.
	
	%Fig. Kr irrad at 650oC, OK.
	
	\subsection{Light-ion implantation at low and high temperatures}
	\label{sec:resdis:lightionimplantation}
	
\begin{figure}[hb!]
		\centering
		\includegraphics[width=0.50\linewidth]{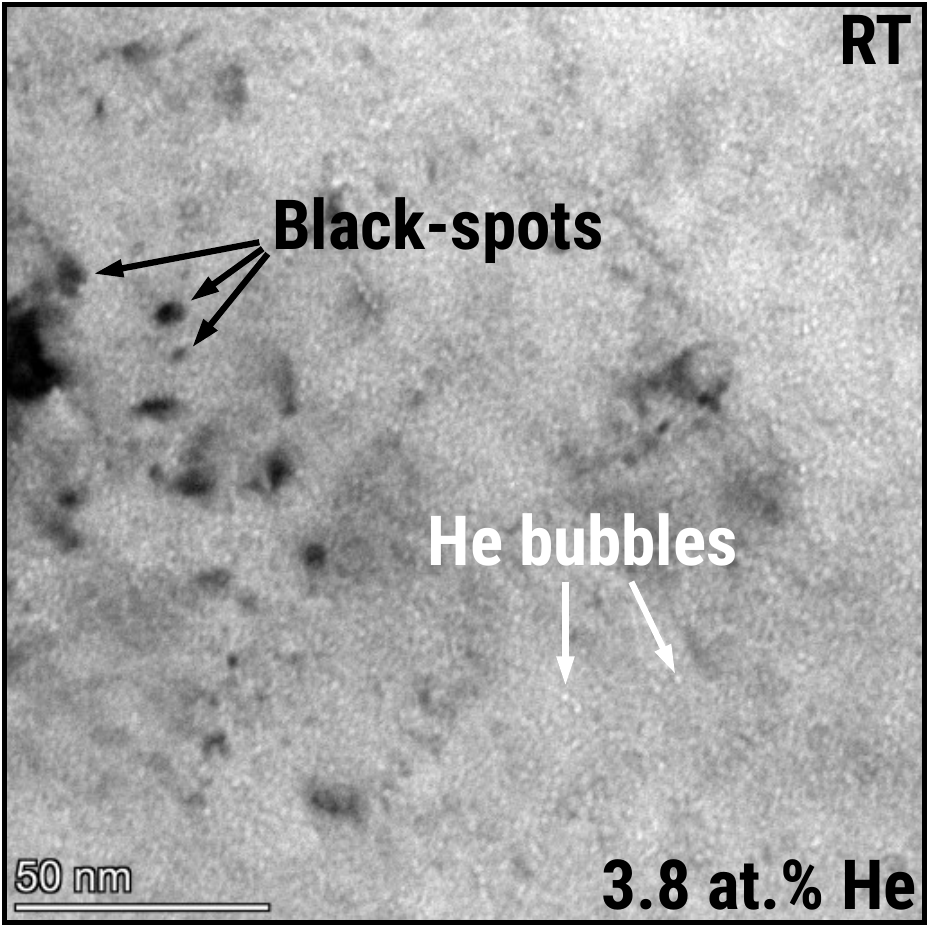}
		\caption{The microstructure of the CuCrZr PA alloy with implanted He at room temperature at an average concentration of 3.80 at\% He. Both He bubbles and black-spots are observed in the microstructure. Note: For the low temperature implantation case, bubbles were only noticed at the end of the experiments; the micrograph was taken with \SI{-2}{\micro\meter} of underfocus.}
		\label{fig:He:RT}
	\end{figure}

\begin{figure}[ht!]
		\centering
		\includegraphics[width=\linewidth]{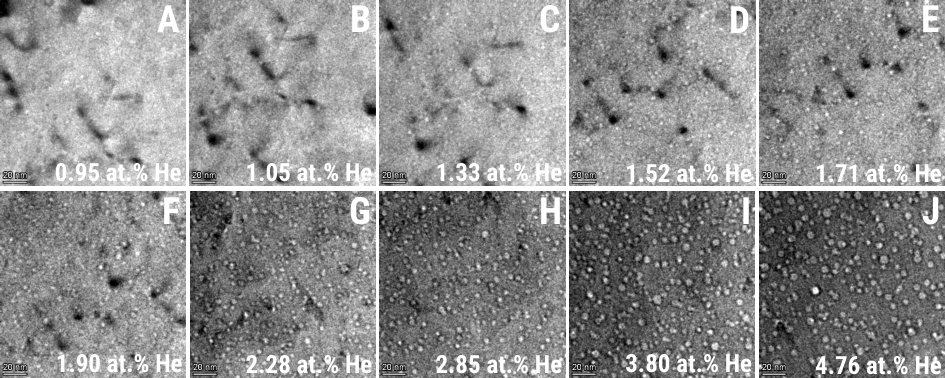}
		\caption{The microstructure of CuCrZr alloy with implanted He at \SI{650}{\degreeCelsius} after heating at this temperature. The BFTEM micrographs figures \textbf{A-J} show the microstructural evolution of the He bubbles from the onset of the observation at 0.95 at\% He until the end of the experiment at 4.76 at\% He. Note: The micrographs were taken with \SI{-2}{\micro\meter} of underfocus.}
		\label{fig:He:650C}
	\end{figure}

	\noindent Transmutation of Cu under fusion neutron irradiation is predicted to generate substantial concentrations of both H and He within the microstructure of the Cu-based heat-sink alloys. Both gas concentrations were estimated at $\sim$\SI{1200}{appm} per year of service in the fusion reactor as calculated by Gilbert \textit{et al.} \cite{gilbert2015handbook}. To assess the microstructural consequences of gas accumulation in the CuCrZr PA microstructure, \SI{6}{\kilo\electronvolt} He$^{+}$ implantations were carried out \textit{in situ} within the TEM at room temperature ($T/T_\mathrm{m} \approx 0.22$) and \SI{650}{\degreeCelsius} ($T/T_\mathrm{m} \approx 0.68$), spanning immobile and mobile defect regimes respectively -- mirroring the thermal conditions of the heavy-ion irradiations in Section~\ref{sec:resdis:heavyionirradiation} -- as shown in Figs.~\ref{fig:He:RT} and~\ref{fig:He:650C}.
	
	%Fig. He implant at RT, OK
	
	At room temperature, He bubbles were not detectable by BFTEM until the implanted concentration reached approximately \SI{3.8}{\atompercent} He, as shown in Fig.~\ref{fig:He:RT}. Below this threshold, the microstructure features black-spot damage clusters analogous to those observed under Kr irradiation, but no resolvable bubble contrast. The delayed onset of bubble visibility points to a sluggish nucleation regime: at $T/T_\mathrm{m} \approx 0.22$, He interstitials have limited thermal mobility within this alloy and must rely on irradiation-induced defect interactions to form stable clusters. Consequently, He accumulates within the matrix -- likely trapped at vacancies and defect clusters generated by the displacement damage events -- until a critical supersaturation is reached at which bubbles become resolvable. This discussion is in-line with state-of-the-art presented by Trinkaus \& Singh, who previously reviewed the behaviour of He in metals and alloys during irradiation \cite{trinkaus2003helium}. The co-existence of black-spot defects and bubbles at \SI{3.80}{\atompercent} He suggests that both displacement damage and He clustering contribute simultaneously to alloy properties' degradation in the low-temperature regime. He atoms implanted at room temperature can trigger self-trapping and trap-mutation events even without pre-existing lattice vacancies: molecular dynamics simulations by Li \textit{et al.} show that a small interstitial He\textsubscript{3} cluster in Cu spontaneously ejects a host atom to form a Frenkel pair, the resulting vacancy capturing the He cluster to form a stable He-vacancy complex \cite{li2024atomistic}. This mechanism, which requires no thermally mobile lattice vacancies, offers a plausible explanation for the elevated He concentration threshold (\SIrange{3.8}{3.9}{\atompercent}) required for TEM-visible bubble nucleation in the present alloy at RT: He-vacancy complexes likely nucleate well below this threshold but remain sub-resolution, with visible bubbles only emerging once sufficient trap-mutation events have accumulated to produce clusters large enough for TEM detection.
	
	%Fig. He implant at 650oC
	
	%Fig. He implant at 650oC coalescence
	\begin{figure}[ht!] 
		\centering 
		\includegraphics[width=\linewidth]{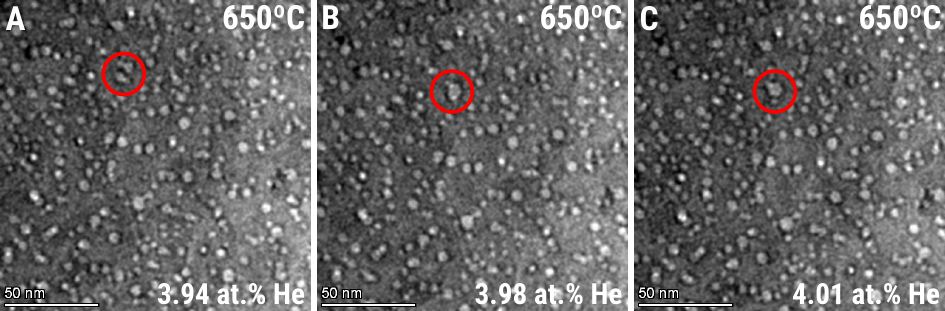} 
		\caption{\textit{In situ} TEM imaging (BFTEM underfocused mode) during \SI{6}{\kilo\electronvolt} He$^{+}$ implantation at \SI{650}{\degreeCelsius} reveals coalescence a the growth mechanism of He bubbles in the CuCrZr alloy with high He concentrations.} 
		\label{fig:He:650C:coalescence} 
	\end{figure}
	
	The implantation behaviour at \SI{650}{\degreeCelsius} is markedly different from the low-temperature case. As shown in the BFTEM sequence of Fig. \ref{fig:He:650C}A--J, bubbles become visible from approximately \SI{0.95}{\atompercent} He -- roughly four times lower than at room temperature -- and their size increases progressively and in a near-linear fashion with concentration up to \SI{4.76}{\atompercent} He. This pronounced reduction in the nucleation threshold (considering the limited spatial resolution of TEMs) arises directly from the enhanced thermal mobility of both vacancies and He interstitials at $T/T_\mathrm{m} \approx 0.68$ \cite{trinkaus2003helium}: mobile vacancies cluster efficiently with implanted He atoms, lowering the supersaturation required for stable bubble nucleation, whilst the subsequent growth proceeds by thermally- and displacive-activated He diffusion and vacancy absorption. The near-linear evolution of bubble size with He concentration is consistent with a growth-dominated regime: nucleation sites are established early and bubbles grow steadily via coalescence as He is supplied by the ion beam, as directly evidenced by \textit{in situ} TEM (Fig.~\ref{fig:He:650C:coalescence}A--C and SupInfo8). 
	
	Whilst first-principles calculations indicate that trapped He atoms stabilise the vacancy and increase its migration barrier relative to a clean vacancy -- rendering individual He-vacancy complexes progressively less mobile as He content increases \cite{gonzalez2014migration} -- the marked enhancement of bubble nucleation at \SI{650}{\celsius} (bubbles visible from as little as \SI{0.95}{\atompercent} He, against \SI{3.8}{}--\SI{3.9}{\atompercent} He at RT) is consistent with growth being driven not by diffusion of the complex itself, but by the much larger thermal vacancy population available at $0.68\,T_{\mathrm{m}}$, which is readily captured by existing He-vacancy embryos and highly mobile interstitial He \cite{li2024atomistic}. In this framework, He-vacancy complexes at \SI{650}{\celsius} grow chiefly via vacancy and He absorption followed by cluster/bubble coalescence (with evidence presented in Fig.~\ref{fig:He:650C:coalescence}A--C and SupInfo8), rather than via long-range migration of the clusters themselves, reconciling the near-linear bubble growth observed at \SI{650}{\celsius} with the comparatively immobile nature of He-loaded vacancies at the atomic scale.
	
	In summary, the He implantation experiments reveal a strong temperature dependence of bubble nucleation and growth in the CuCrZr alloy. At room temperature, the high supersaturation required for bubble nucleation and the simultaneous accumulation of displacement damage suggest that He embrittlement will be exacerbated at low operating temperatures. At higher temperatures, the early onset and progressive growth of bubbles at comparatively low He concentrations indicate that thermally activated processes further accelerate microstructural degradation, with potential consequences for the mechanical integrity of the heat-sink alloy over extended service lifetimes.
	
	%Fig. Pristine and Kr irrad at both RT and 650oC STEM-EDX, OK.
	\begin{figure}[ht!]
		\centering
		\includegraphics[width=\linewidth]{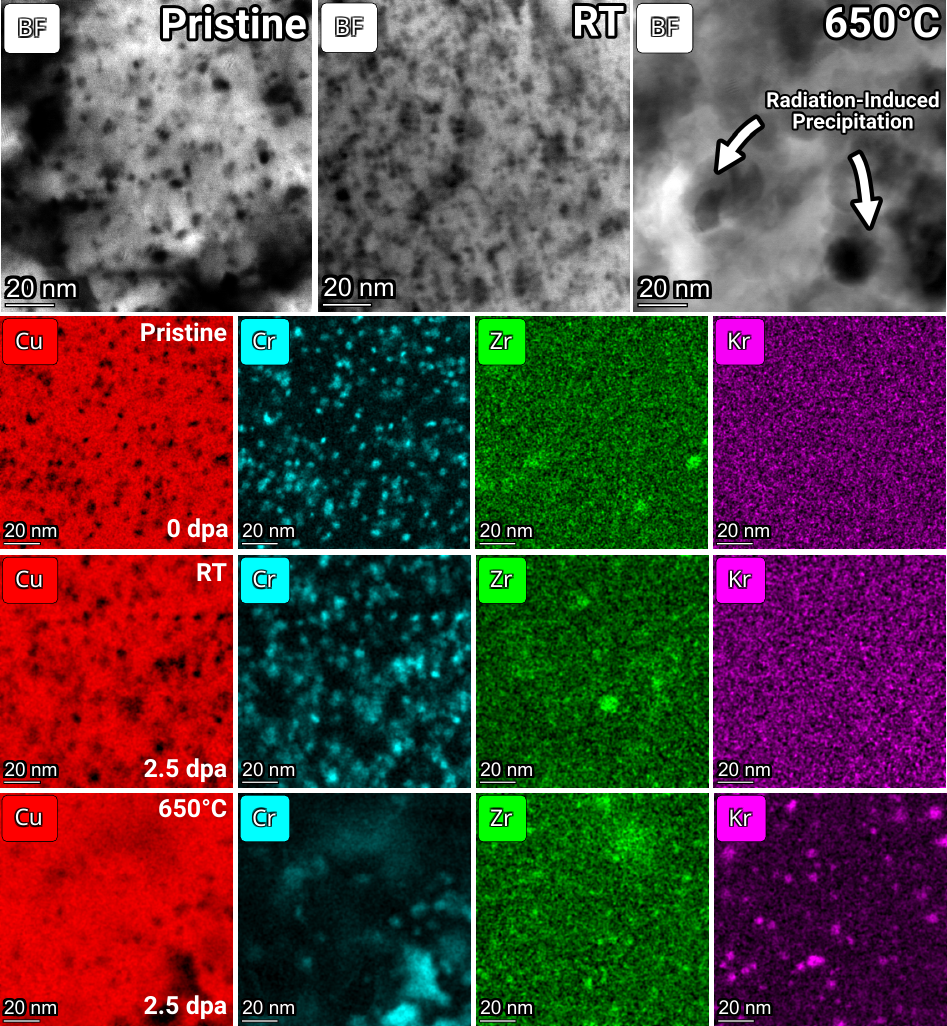}
		\caption{A set of BF-STEM and STEM-EDX maps showing the thermodynamic state evolution of the CuCrZr PA alloy microstructure at different temperatures (RT and \SI{650}{\celsius}) and different doses (0 and 2.5 dpa).}
		\label{fig:ChemicalMaps}
	\end{figure}
	
	\subsection{Pre- and post-analytical characterisation and quantification of defects and precipitates}
	\label{sec:resdis:quantification}
	
	\noindent Having qualitatively described the microstructural evolution of precipitates and irradiation-induced microstructural defects such as black-spots and inert gas bubbles in subsections \ref{sec:resdis:alloyPAasreceived}, \ref{sec:resdis:heavyionirradiation} and \ref{sec:resdis:lightionimplantation}, we now turn to analytical STEM-EDX characterisation and quantitative size statistics to substantiate these observations. This combines chemical mapping of the pristine (PA state unirradiated) and Kr-irradiated conditions at RT and \SI{650}{\celsius} with quantification of Cr- and Zr-rich precipitate diameters, He bubble volume as a function of implanted concentration. 
	
	%Fig. Histograms for precipitate and bubble sizes together with plot of bubble size versus irradiation time/dose.
	\begin{figure}[ht!]
		\centering
		\includegraphics[width=1.0\linewidth]{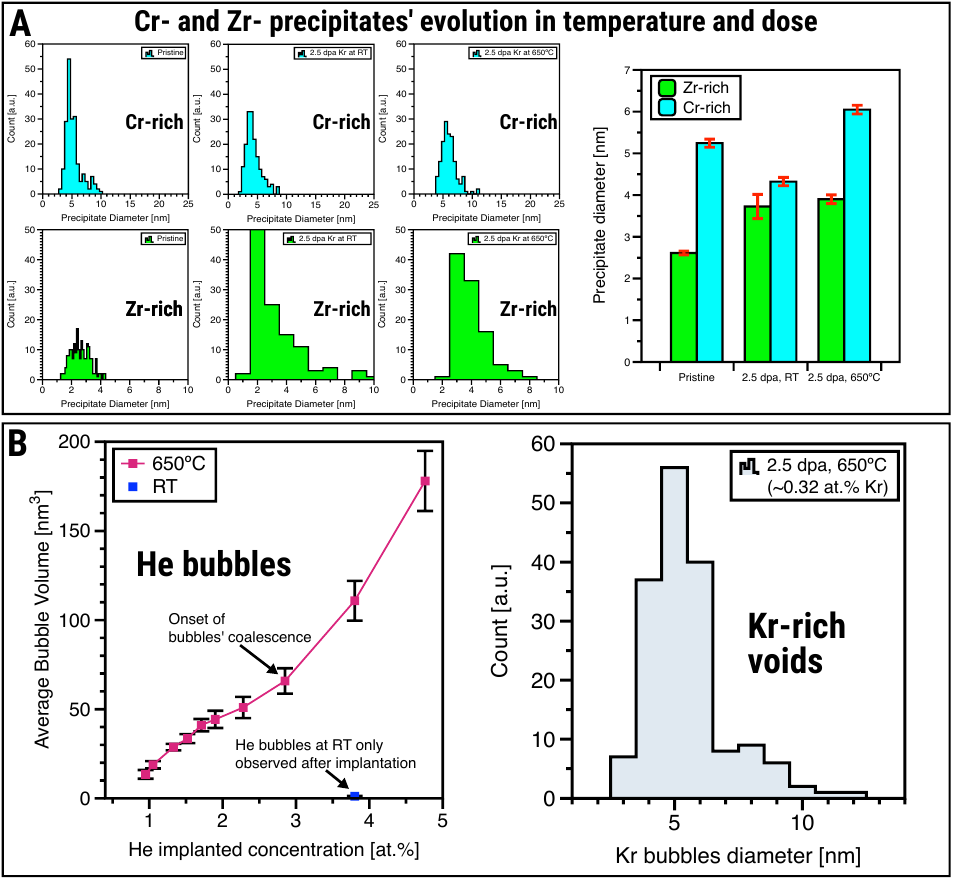}
		\caption{Quantification of changes in the PA heat-treatment state as a result of both heavy-ion irradiations and light-ion implantations. \textbf{A} The evolution of Cr- and Zr-rich precipitate sizes as a function of temperature and dose. \textbf{B} Average bubble volume as a function of implanted He concentration at both RT and \SI{650}{\celsius}, and the Kr-rich voids size histogram at \SI{650}{\celsius} with approximately \SI{0.32}{\atompercent} implanted Kr as a consequence of the heavy-ion irradiation experiments.}
		\label{fig:bubbles:precipitates:histogram}
	\end{figure}
	
	Fig. \ref{fig:ChemicalMaps} shows BF-STEM and STEM-EDX maps of the alloy in the pristine, 2.5 dpa RT and 2.5 dpa \SI{650}{\celsius} conditions. In the PA and unirradiated state, the Cr-rich and Zr-rich nanoprecipitates described in Section \ref{sec:resdis:alloyPAasreceived} are clearly resolved in the nanoscale EDX maps. After irradiation to 2.5 dpa at RT, both precipitate populations appear to fade and lose their characteristic rounded-shape in the chemical maps, which is (potentially) consistent with the ballistic dissolution discussed in Section \ref{sec:resdis:heavyionirradiation}, although amorphisation of the nano-precipitates cannot also be ruled out. At 2.5 dpa and \SI{650}{\celsius}, the original nano-precipitate population is no longer observed. Instead, a coarser population of Cr-rich precipitates together with a finer population of Zr-rich precipitates is detected, a redistribution we attribute to RIP rather than to survival of the as-received precipitates' population achieved after PA heat-treatment. A new observation brought by the STEM-EDX mapping post-irradiation characterisation was the presence of voids filled with Kr after high-temperature irradiation. 
	
	The precipitate size statistics in Fig. \ref{fig:bubbles:precipitates:histogram}A quantify the microstructural evolution of the CuCrZr PA alloy during to both heavy- and light-ion irradiations/implantations. In the pristine condition, the Cr-rich precipitates show a narrow size distribution centred at approximately \SI{5.3}{nm}, while the Zr-rich precipitates are smaller, with a mean diameter of approximately \SI{2.6}{nm}, consistent with the two distinct precipitate variants identified by STEM-EDX and SAED in subsection \ref{sec:resdis:alloyPAasreceived}. After irradiation to 2.5 dpa at RT, the mean Cr-rich precipitate diameter decreases slightly to approximately \SI{4.3}{nm} with a broader and more dispersed (faded) distribution, while the Zr-rich population coarsens to approximately \SI{3.8}{nm} on average and develops a long tail extending beyond \SI{8}{nm}. At 2.5 dpa and \SI{650}{\celsius}, the Cr-rich precipitates that arose due to RIP are approximately \SI{6.1}{nm} in size, while the Zr-rich precipitates stabilise at approximately \SI{3.9}{nm}, only marginally larger than in the RT-irradiated condition. This combination of an enlarged, broadened Cr-rich population and a comparatively narrower Zr-rich population at high temperature is consistent with thermally activated coarsening operating alongside radiation-accelerated diffusion, although the relative contribution of each mechanism cannot be isolated from precipitate size statistics alone.
	
	Fig. \ref{fig:bubbles:precipitates:histogram}B quantifies average He bubble volume as a function of He concentration, as introduced in subsections \ref{sec:resdis:heavyionirradiation} and \ref{sec:resdis:lightionimplantation}. Beyond \SI{3}{\atompercent} He, bubble volume grows linearly, indicating that coalescence has become the dominant coarsening mechanism. The higher nucleation threshold at RT ($T_m = 0.22$) compared with \SI{650}{\celsius} ($T_m = 0.68$) suggests the limited thermal mobility of He and vacancies at low temperature, which could confine bubble evolution to a nucleation-dominated regime rather than allowing coalescence-driven growth. For the heavy-ion case, the Kr-rich void size distribution at 2.5 dpa and \SI{650}{\celsius}, corresponding to approximately \SI{0.32}{\atompercent} implanted Kr, peaks between \SIrange{4}{6}{\nano\metre} with a tail extending to approximately \SI{12}{\nano\metre}.
	
	\section{Overall Discussion}
	\label{sec:overall}
	
	\subsection{Equilibrium assessment after both prime-ageing and within a thermonuclear fusion reactor}
	\label{sec:overall:equilibrium}
	
	\noindent The sections of the Cu-rich corner of the Cu-Cr-Zr ternary system at \SI{475}{\degreeCelsius} and \SI{600}{\degreeCelsius} indicate that the as-received CuCuZr alloy samples with 0.67 wt.\% Cr and 0.083 wt.\% Zr comprise the following solution phases:
	
	\begin{itemize}
		\item At \SI{475}{\degreeCelsius}, the PA heat-treatment: (i) FCC\_A1, (ii) BCC\_A2, and (iii) LAVES\_C15; and
		\item At \SI{600}{\degreeCelsius}, the high-temperature irradiation/implantation: (i) FCC\_A1, (ii) BCC\_A2, and (iii) Cu$_5$Zr.
	\end{itemize}
	
	\begin{figure}[ht!]
		\centering
		\includegraphics[width=1.0\linewidth]{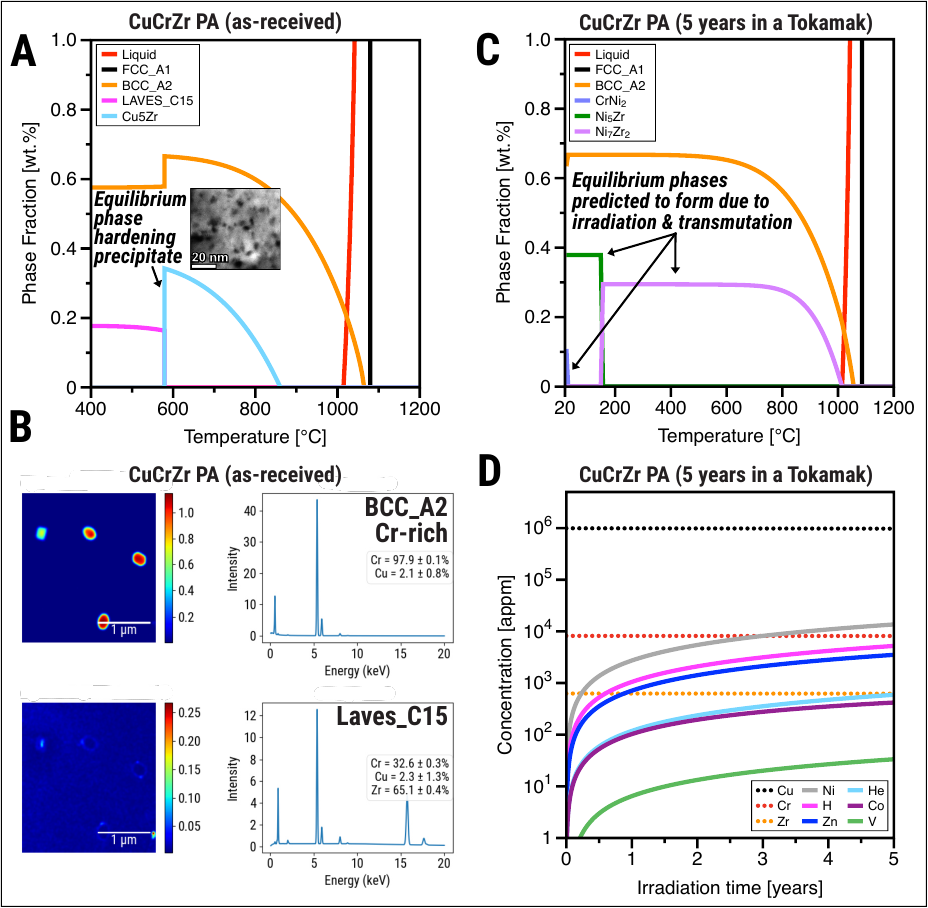}
		\caption{Thermodynamic and transmutation calculations performed with FactSage and FISPACT-II. \textbf{A} The equilibrium phase diagram as a function of temperature obtained for the as-received CuCrZr PA alloy; the BFTEM inset shows the hardening nano-precipitates observed in the unirradiated PA microstructure. \textbf{B} ESPM decomposition of the STEM-EDX maps collected from the CuCrZr PA alloy, confirming the presence of both the Cr-rich BCC\_A2 and Laves\_C15 phases in the alloy microstructure. \textbf{C} The equilibrium phase diagram recalculated for the bulk alloy composition after five full-power years of service in a fusion reactor, incorporating the transmutation products listed in Table~\ref{tab:transmutation}. \textbf{D} Evolution of the transmuted elemental concentrations of the CuCrZr alloy over five years of irradiation time.}
		\label{fig:Equilibrium}
	\end{figure}

	This equilibrium is better understood by considering the cooling calculations for the as-received CuCrZr alloy samples, as shown in Fig.~\ref{fig:Equilibrium}A. The solidification of the melt starts at about \SI{1080}{\degreeCelsius} with the formation of large amounts of FCC\_A1 (matrix), while the precipitation of small amounts of BCC\_A2 starts at \SI{1060}{\degreeCelsius}, and that of Cu$_5$Zr at around \SI{870}{\degreeCelsius}. On continued cooling, the intermetallic compound Cu$_5$Zr reacts with some BCC\_A2 at \SI{578}{\degreeCelsius} to form LAVES\_C15 and FCC\_A1, a peritectoid reaction. Low-magnification STEM-EDX mapping together with the ESPM code allowed us to identify both the Cr-rich BCC\_A2 and Laves\_C15 phases at the microscale, as shown in Fig.~\ref{fig:Equilibrium}B. 
	
	Neither of these two equilibrium phases, however, corresponds directly to the nanoscale hardening precipitates identified in subsection \ref{sec:resdis:alloyPAasreceived}, in agreement with the metastability of CuCrZr alloys previously investigated with TEM and APT by Hughes \textit{et al.} \cite{hughes2022fullstage}. A Cr-rich nano-precipitate has also been observed in the microstructure of the CuCrZr PA as-received alloy and is shown in Fig. \ref{fig:Pristine}E, possibly hinting at the BCC\_A2 phase (Cr-rich with dissolved Zr) forming at this finer length scale. By contrast, the Zr-rich nanoprecipitates were not predicted at all by the equilibrium calculations, despite being clearly resolved in the same STEM-EDX assessment in Fig. \ref{fig:Pristine}E. This apparent discrepancy between the database and the characterisation experiments is expected: precipitation during ageing is an inherently non-equilibrium process, and the observed metastable dispersion is kinetically favoured over the equilibrium phases predicted by FactSage. Alternatively, it may indicate that the hardening precipitates are not simply a finely divided form of the bulk equilibrium phases, but rather a distinct, kinetically stabilised population that forms during the PA heat-treatment and is not itself captured by an equilibrium assessment of the Cu-Cr-Zr system. Resolving this would require either an extended thermodynamic database incorporating these metastable Cu-Zr phases, or a dedicated kinetic simulation (\textit{e.g.}, Dictra) of the PA heat-treatment itself. It is worth emphasising that the ternary phase diagrams and the respective calculated compositions of the phases detected in the CuCrZr PA alloy are shown in the supplemental information (SupInfo5).
	
	This discussion above, however, only describes the alloy as it leaves the manufacturing line. Table \ref{tab:transmutation} and Fig. \ref{fig:Equilibrium}D quantify how the bulk composition of the CuCrZr alloy is predicted to evolve under direct fusion neutron irradiation as calculated by us using FISPACT and FactSage: after five full-power years of service, transmutation introduces approximately \SI{1.35}{\atompercent} Ni, \SI{0.52}{\atompercent} H, \SI{0.35}{\atompercent} Zn, together with smaller additions of Co, V and Ti, on top of an $\sim$0.8~at.\% increase in Cr and a more modest increase in Zr arising from transmutation of Cu itself (the SupInfo2, 3, and 4 shows the full evolution of the alloy under fusion neutrons). Recalculating the Cu--Cr--Zr--Ni--Zn (not considering the minor Co, VTi) equilibrium for this transmuted composition reveals that the phase landscape of the alloy is fundamentally altered when compared with the initial CuCrZr alloy. These results are shown in Fig. \ref{fig:Equilibrium}C: while the Cr-rich BCC\_A2 phase persists across essentially the same temperature range as in the as-fabricated alloy, all the other original phases of the Cu--Cr--Zr system are entirely absent from the post-transmutation equilibrium. In their place, transmutation-derived Ni reacts preferentially with Zr and Cr to stabilise a new set of intermetallics -- CrNi$_2$ and Ni$_5$Zr at low temperature, and Ni$_7$Zr$_2$ across a broad temperature window extending from below \SI{150}{\degreeCelsius} up to nearly \SI{900}{\degreeCelsius}, comfortably spanning the ITER operating window of \SIrange{200}{350}{\degreeCelsius} \cite{hirai2016use}. In other words, the Zr that the alloy was designed to partition into Cu-based, coherent hardening precipitates is instead redirected by transmutation towards a Ni-bearing intermetallic system that bears no resemblance, in chemistry, morphology or coherency, to the nanoscale Cr- and Zr-rich precipitates characterised in subsection \ref{sec:resdis:alloyPAasreceived}.
	
	This compositional drift is mechanistically distinct from, and additive to, the displacement-damage-driven precipitate dissolution and radiation-induced precipitation (RIP) to be discussed next in subsection \ref{sec:overall:precipitates}. The latter concerns the fate of the existing PA precipitates' population under ballistic mixing and thermally activated diffusion. The former concerns the fact that, even in the absence of any displacement damage whatsoever, the thermodynamically favoured precipitate chemistry of the alloy changes irreversibly as transmutation proceeds. Within five years of service -- a small fraction of the multi-decade design lifetime anticipated for a commercial fusion reactor \cite{EUROfusion2018Roadmap,UKFusionMaterialsRoadmap2040} -- the alloy is predicted to no longer be the Cu-Cr-Zr system it was designed and heat-treated as, but rather a Cu--Cr--Zr--Ni system whose equilibrium precipitate chemistry has not been characterised, optimised, or validated for heat-sink performance. In addition, it is historically known that the presence of Ni is deleterious for nuclear materials in general as it is prone to transmute to He at large rates, forming nano-bubbles that contribute to severe embrittlement \cite{maziasz1985helium}. 
	
	Taken together, the equilibrium assessment presented here shows that the CuCrZr alloy and its age-hardening strategy is compromised on two independent and compounding fronts: 
	
	\begin{itemize}
		\item[(i)] The as-fabricated precipitates' population at nanoscale is itself a metastable, far-from-equilibrium state that is (or can be) easily destabilised by energetic particle irradiation and thermal excursions such as overheating, or even a combination thereof; while 
		
		\item[(ii)] The underlying alloy chemistry on which the entire precipitation-hardening design is predicated is progressively and irreversibly altered by nuclear transmutation over reactor service life, creating a new metastable equilibrium state that may compromise the starting mechanical and thermal properties of the alloy.
	\end{itemize}
	
	This latter effect is not specific to the displacement-damage mechanisms explored experimentally in this work, and cannot be mitigated by alternative heat-treatments or precipitate engineering within the existing Cu--Cr--Zr alloy system. It follows that reliance on a precipitation-hardened, age-hardenable Cu-based alloy is not an appropriate long-term strategy for fusion reactor heat-sink applications: the very transmutation reactions that are unavoidable under a 14~MeV fusion neutron spectrum ensure that the alloy's thermodynamic metastable state, and with it the chemical basis for precipitation strengthening, will not remain fixed over the component's service life.
	
	\begin{table}
		\centering
		\caption{Transmutation products from the CuCrZr alloy after five full power years in a fusion reactor environment.}
		\label{tab:transmutation}
		\begin{tabularx}{\linewidth}{l X X X}
			\toprule
			\textbf{Element} & \centering\textbf{No.\ of Atoms [\#]} & \centering\textbf{Concentration [appm]} & \centering\arraybackslash\textbf{Concentration [at.\%]} \\
			\midrule
			Cu & \centering $9.24 \times 10^{24}$ & \centering $9.68 \times 10^{5}$ & \centering\arraybackslash 96.803 \\
			Ni & \centering $1.29 \times 10^{23}$ & \centering $1.35 \times 10^{4}$ & \centering\arraybackslash  1.353 \\
			Cr & \centering $7.72 \times 10^{22}$ & \centering $8.09 \times 10^{3}$ & \centering\arraybackslash  0.809 \\
			H  & \centering $4.96 \times 10^{22}$ & \centering $5.19 \times 10^{3}$ & \centering\arraybackslash  0.519 \\
			Zn & \centering $3.33 \times 10^{22}$ & \centering $3.49 \times 10^{3}$ & \centering\arraybackslash  0.349 \\
			Zr & \centering $5.91 \times 10^{21}$ & \centering $6.19 \times 10^{2}$ & \centering\arraybackslash  0.062 \\
			He & \centering $5.61 \times 10^{21}$ & \centering $5.88 \times 10^{2}$ & \centering\arraybackslash  0.059 \\
			Co & \centering $3.97 \times 10^{21}$ & \centering $4.16 \times 10^{2}$ & \centering\arraybackslash  0.042 \\
			V  & \centering $3.17 \times 10^{20}$ & \centering $3.32 \times 10^{1}$ & \centering\arraybackslash  0.003 \\
			Ti & \centering $5.54 \times 10^{19}$ & \centering $5.81 \times 10^{0}$ & \centering\arraybackslash  0.001 \\
			\bottomrule
		\end{tabularx}
		\vspace{2pt}
		\begin{flushleft}
			\footnotesize$^{1}$ Elements are ranked by atomic concentration. \\[3pt]
			\footnotesize$^{2}$ Only parent and major daughter nuclei above 0.001 at\% are listed. \\[3pt]
			\footnotesize$^{3}$ Calculations were performed considering 1 kg of the CuCrZr alloy.
		\end{flushleft}
	\end{table}
	
	\subsection{Precipitate metastability limits age-hardenable Cu-based alloys as thermonuclear fusion reactor heat-sinks}
	\label{sec:overall:precipitates}
	
	\noindent The materials selection criteria for thermonuclear fusion reactors as outlined in section~\ref{sec:introduction} states that a heat-sink alloy must simultaneously deliver a trinity of high thermal conductivity, high strength and high irradiation resistance. The CuCrZr PA alloy satisfies the first two criteria through a precipitation-hardening route that depends on a fine, homogeneously dispersed population of Cr- and Zr-rich nano-precipitates \cite{bochvar2007cr,edwards2007effect,fabritsiev2005effect,kalinin2007ageing}, the existence of which was directly confirmed in the PA condition by STEM-EDX, SAED and HRTEM in subsection~\ref{sec:resdis:alloyPAasreceived}. The present results show that this precipitate population, far from being the robust microstructural feature implied by its role as the principal strengthening mechanism, is in fact doubly unstable: it is metastable with respect to temperature alone, as for example when the alloy in PA state is heated up to \SI{600}{\celsius} as shown in Fig. \ref{fig:Kr:650C}A,E, and independently destabilised by displacement damage, through two distinct but equally deleterious pathways depending on the irradiation temperature: one in which defects are immobile and a higher temperature one, where defects are mobile.
	
	The thermal instability of the PA precipitate population is established on two independent grounds. First, the hardness experiments in subsection \ref{sec:resdis:alloyPAasreceived} demonstrate that solution annealing at \SI{980}{\celsius} followed by re-ageing at \SI{475}{\celsius} does not restore the as-received PA hardness (140 HBW), recovering only 65 HBW after SA and an intermediate value thereafter -- evidence that a fraction of the original strength originates from plastic deformation and dislocation substructures rather than from the precipitates alone, consistent with observations from the previous work of Gong \textit{et al.} \cite{gong2024electropolishing_CuCrZr}. Second, and more fundamentally, the thermodynamic assessment in subsection \ref{sec:overall:equilibrium} shows that the fine, coherent Cr- and Zr-rich nano-precipitate dispersion observed after PA is not the true equilibrium product at fusion-relevant temperatures, especially under overheating conditions. The PA nano-precipitate population is therefore a kinetically trapped, far-from-equilibrium state, and any thermal excursion above the ageing temperature -- whether from reactor transients or from the elevated temperatures explored in this work -- drives the microstructure towards a coarser, less effective equilibrium configuration, compromising the required high-strength levels for the heat-sink pipes. This is precisely the behaviour reported by Kalinin \textit{et al.} \cite{kalinin2007ageing}, in which only one hour of overheating at \SI{600}{\degreeCelsius} reduces the yield strength of PA CuCrZr to the level of the solution-annealed condition (Fig. \ref{fig:LitRev}A).
	
	Energetic particle irradiation exacerbates this intrinsic thermal-induced degradation of mechanical properties rather than acting as an independent degradation mode. At room temperature, the \SI{600}{\kilo\electronvolt} Kr$^{2+}$ irradiation experiments shown in subsection \ref{sec:resdis:heavyionirradiation} suggest that the superlattice diffraction reflections associated with the hardening precipitates progressively weaken and vanish with dose. Additionally, the STEM-EDX maps of subsection \ref{sec:resdis:quantification} indicate a loss of the rounded-shape morphology and a decrease in mean Cr-rich precipitate diameter -- evidence of ballistic dissolution of the precipitates concurrently with black-spot defect accumulation in the matrix. For the irradiations at \SI{650}{\degreeCelsius}, the precipitates are destabilised by a different route: thermally activated dissolution is followed by radiation-induced precipitation (RIP), yielding a coarser (a new) Cr-rich phase population (\SI{6.1}{nm}, the largest mean diameter recorded across all conditions examined) and a redistributed Zr-rich population, rather than survival of the original PA nano-precipitate microstructure. In both thermal regimes, therefore, the metastable nanoscale hardening precipitates that defines the PA strengthening condition is completely destroyed, either by ballistic disordering at low homologous temperature or by a combination of thermally- and radiation-driven coarsening at high homologous temperature: there is no irradiation/temperature window within the range examined here in which the original precipitate microstructure survives intact. Interestingly, these results are in agreement with experiments performed on age-hardenable aluminium alloys, where energetic particle irradiation is also reported to destroy and dissolve nano-precipitates, compromising the aluminium alloys' designed mechanical properties \cite{lohmann1987amicrostructure,tunes2026legacy}.
	
	The findings above have a direct bearing on the apparent mechanical resilience reported for irradiated CuCrZr PA alloy in the literature. Fabritsiev \& Pokrovsky \cite{fabritsiev2005effect} and Edwards \textit{et al.} \cite{edwards2007effect} independently reported that neutron irradiation increases the yield strength of PA CuCrZr at low dose and low-to-moderate temperature, but this is accompanied by a severe loss of ductility -- uniform elongation falling from 25\% to 10\% after only 0.54 dpa at \SI{150}{\degreeCelsius}. The present microstructural evidence unravelled by our work offers a mechanistic explanation for this trade-off: the apparent ``hardening'' is not a continuation of precipitation strengthening, but a substitution of one strengthening mechanism for another. As the engineered precipitate dispersion dissolves under irradiation, strength is instead sustained by the accumulating population of unresolved, irradiation-induced black-spot defect clusters identified in subsection \ref{sec:resdis:heavyionirradiation} when heavy-ion irradiations mimic the neutron spectrum of Cu PKA in a typical fusion reactor. This defect-cluster hardening is mechanistically equivalent to the radiation hardening and associated embrittlement well documented in irradiated metals generally, and is fundamentally distinct from the ductile, controlled precipitation hardening that the PA heat-treatment was designed to provide. The CuCrZr PA alloy therefore loses the very microstructural feature responsible for its favourable strength--ductility balance precisely under the conditions it is intended to withstand (\textit{i.e.}, energetic particle irradiation in a regime where defects are mobile).
	
	Taken together, these observations raise questions about whether an age-hardenable Cu-based alloy is well suited to this application. The strengthening mechanism relies on a metastable nano-precipitate population that is unstable with respect to temperature alone, is further destabilised by displacement damage via two distinct but equally damaging pathways depending on operating temperature, and is ultimately replaced -- not preserved -- by an uncontrolled, embrittling and defective microstructure once irradiation begins. As heat-sink components must simultaneously tolerate both elevated operating and transient temperatures \cite{hirai2016use} and sustained displacement damage accumulating over reactor lifetime \cite{EUROfusion2018Roadmap,UKFusionMaterialsRoadmap2040}, a strengthening route that is sensitive to either factor individually -- and, as shown in this present work, to both simultaneously in this alloy -- may struggle to provide a stable, predictable mechanical and thermal responses over the service life of a fusion reactor heat-sink. This adds a precipitate-stability dimension to the broader limitations of Cu-based alloys for fusion applications already noted in the previous subsection \ref{sec:overall:equilibrium} and points towards the need for Cu-based alloy heat-sink strengthening strategies that do not rely on a thermodynamically metastable precipitate dispersion as their primary strengthening microstructural feature.
	
	\revision{One last pertinent question is whether the precipitate instabilities herein observed between room temperature and \SI{650}{\celsius} permits extrapolation to the in-service window of \SIrange{200}{350}{\celsius} for ITER and up to \SI{450}{\celsius} envisaged for DEMO. The two irradiation temperatures employed here should be regarded as end-members of a continuous kinetic spectrum rather than as isolated conditions: at room temperature (\num{0.22}\,$T_\mathrm{m}$), vacancy mobility is suppressed and ballistic dissolution proceeds unopposed, whereas at \SI{650}{\celsius} (\num{0.68}\,$T_\mathrm{m}$), thermally-activated dissolution and re-precipitation dominate. The service window corresponds to homologous temperatures of approximately \numrange{0.46}{0.53}\,$T_\mathrm{m}$, at which vacancies in Cu are already mobile; consequently, both ballistic mixing and thermally-assisted solute redistribution are expected to operate concurrently in service. Since the prime-aged dispersion degrades at each kinetic extreme individually, no intermediate temperature window exists in which one mechanism could be suppressed without the other becoming dominant, and the bracketing conditions studied here therefore bound, rather than bypass, the in-service behaviour.}
	
	\subsection{Effect of gaseous transmutation products on the CuCrZr alloy microstructure}
	\label{sec:overall:transmutation}
	
	\noindent Irradiation resistance in a fusion-relevant neutron spectrum cannot be assessed in terms of displacement damage alone. As shown in Table~\ref{tab:transmutation} and Fig.~\ref{fig:Equilibrium}D, transmutation of Cu generates substantial H and He, reaching approximately \SI{0.52}{\atompercent} H and \SI{0.06}{\atompercent} He (not considering the progressive transmutation of Ni) after five full-power years \cite{gilbert2015handbook}. Unlike the solid transmutation products discussed in subsection \ref{sec:overall:equilibrium}, H and He are essentially insoluble in Cu and instead accumulate as gas, directly affecting cavity nucleation and growth.
	
	The light-ion implantation experiments in subsection \ref{sec:resdis:lightionimplantation} provide a mechanistic basis for interpreting this accumulation. At room temperature, He bubbles were only resolved above \SI{3.8}{\atompercent} He, reflecting sluggish nucleation at $T/T_\mathrm{m} \approx 0.22$. At \SI{650}{\celsius}, bubbles appeared from approximately \SI{0.95}{\atompercent} He and grew steadily, reaching \SI{7.0}{\nano\metre} at \SI{4.76}{\atompercent} He. The five-year reactor-predicted concentration sits below both thresholds; given the much higher generation rates used in this \textit{in situ} study relative to reactor service, this should not be read as a literal prediction of onset time, but it confirms that the nucleation threshold at elevated, fusion-relevant temperature is modest and plausibly attainable over a multi-decade reactor lifetime \cite{EUROfusion2018Roadmap,UKFusionMaterialsRoadmap2040}.
	
	\revision{A comparison with the Kr-rich cavity population formed at \SI{650}{\celsius} (subsection \ref{sec:resdis:quantification}) sharpens this picture, but also raises a question: are these cavities voids that would nucleate regardless of gas content, with Kr simply partitioning into them because it was implanted, or does the insoluble gas actively promote their formation? The present dataset cannot distinguish the two. Vacancies at \SI{650}{\celsius} are mobile enough to agglomerate into voids unassisted, so the observed cavities may simply reflect the irradiation temperature rather than any Kr-specific effect. Yet despite the implanted Kr concentration being roughly an order of magnitude lower than the He concentrations explored at the same temperature, the resulting Kr-rich voids reach comparable sizes to He bubbles formed at around \SI{1}{\atompercent} He -- a disproportionality more consistent with the insoluble gas stabilising cavities against thermal re-dissolution, or assisting nucleation outright, than with it merely decorating pre-existing voids. Regardless of which contribution dominates, the vacancy mobility that drives cavity growth at \SI{650}{\celsius} is a condition equally conducive to void swelling under neutron irradiation, where insoluble transmutant gases are likewise present.}
	
	In our study, H was not significantly implanted, but Table \ref{tab:transmutation} shows it is predicted to accumulate at roughly an order of magnitude higher concentration than He over the same period. H interacts strongly with vacancies, dislocations and grain boundaries and is notoriously difficult to characterise directly in metallic microstructures \cite{tunes2024limitations}. The combined presence of H and He at these concentrations raises the possibility of synergistic, mixed-species cavity nucleation, as documented in other multi-species irradiation environments \cite{ullmaier1985He,marian2015synnergyHandHe,dias2017synnergyHandHe}. This expectation has, in fact, recently been confirmed directly in CuCrZr(Y) alloys by Zhang \textit{et al.} \cite{zhang2026precipitate}, who showed via triple-beam (H$^{+}$ + He$^{+}$ + Fe$^{2+}$) irradiation at \SI{450}{\degreeCelsius} that measurable swelling only develops in the region co-implanted with H and He, whereas an equivalent region subjected to Fe$^{2+}$ displacement damage alone accumulates comparable defect densities but no resolvable cavities -- a direct demonstration that gas co-implantation, rather than displacement damage in isolation, drives cavity nucleation in this alloy system. Notably, in that same study, cavities preferentially nucleated at the interfaces of coarsened Cr-rich precipitates and Zr-rich clusters, reinforcing the link between precipitate instability (discussed herein in subsection~\ref{sec:overall:precipitates}) and gas-driven cavity formation discovered in this present work. 
	
	Still on the topic of transmuted H in the CuCrZr alloy, it is worth emphasising that in a recent study, we have demonstrated that bubbles and even nano-cracks can form in the microstructure of a martensitic steel when implanted with H \cite{borges2026lowenergy}, so one cannot rule out the possibility of severe synergistic H-induced embrittlement \cite{djukic2016hydrogenembrittlement, djukic2015hydrogendamage, djukic2019synergistic, koyama2017recent} in Cu-based alloys whilst in-service in fusion reactors. In Cu alloys, this risk is compounded by a purely chemical degradation route known since the earliest studies of ``hydrogen disease'' in non-ferrous metallurgy: any residual oxygen in the alloy -- typically present as Cu$_2$O inclusions segregated at grain boundaries -- is reduced by the presence of H to form H$_2$O vapour (Cu$_2$O + 2H $\rightarrow$ 2Cu + H$_2$O), and the extreme volume expansion accompanying internal steam formation generates pressures sufficient to nucleate grain-boundary cavities and intergranular cracks \cite{wampler1976hydrogen,nieh1980water}. Under the H generation rates predicted in Table~\ref{tab:transmutation}, even trace oxygen contamination could therefore act as a potent embrittlement vector, and the use of strictly oxygen-free Cu feedstock -- as already specified for ITER-grade CuCrZr -- must be regarded as a prerequisite for H tolerance in fusion-relevant Cu alloys.
	
	Taken together with subsections~\ref{sec:overall:equilibrium} and \ref{sec:overall:precipitates}, the results in this paper indicate that gaseous transmutation products constitute a third, largely independent degradation pathway for the CuCrZr PA alloy: one operating through cavity nucleation and growth rather than precipitate loss or compositional drift, but equally intrinsic to Cu behaviour under a fusion neutron spectrum. 
	
	\section{Conclusions}
	\label{sec:conclusions}
	\noindent The CuCrZr heat-sink alloy was characterised in the PA condition and subjected to \textit{in situ} heavy-ion irradiation, light-ion implantation, analytical STEM-EDX quantification pre- and post-irradiation, and thermodynamic/transmutation modelling. The following conclusions can be drawn:
	
	\begin{enumerate}
		\item The as-received PA alloy combines a fine, equiaxed grain structure -- average size of \SI{9.38(26)}{\micro\metre} -- with two coexisting nano-precipitate variants, Cr-rich (\SI{5.3}{\nano\metre}) and Zr-rich (\SI{2.6}{\nano\metre}). PA hardness (140~HBW) is only partially recovered after re-ageing (120 HBW), showing it arises from precipitation and retained plastic deformation during processing, not precipitation alone.
		
		\item Heavy-ion irradiation destabilises the hardening precipitates via two distinct routes: radiation-induced dissolution at room temperature (Cr-rich diameter falling to \SI{4.3}{\nano\metre}), and thermally- and radiation-driven dissolution followed by radiation-induced precipitation at \SI{650}{\degreeCelsius} (coarsened Cr-rich phase, \SI{6.1}{\nano\metre}). The original PA microstructure survives in neither regime; as dissolution occurs under opposing kinetic conditions, it cannot be attributed to a dose-rate artefact of either. Precipitate amorphisation cannot be ruled out.
		
		\item Cavity nucleation is strongly temperature-dependent with the He bubbles observation threshold falling from \SIrange{3.8}{3.9}{\atompercent} at room temperature to \SI{0.95}{\atompercent} at \SI{650}{\degreeCelsius}, where bubbles grow by coalescence to \SI{7.0}{\nano\metre}. On the other hand, Kr-rich voids reach comparable sizes at an order of magnitude lower concentration, indicating that cavity formation is governed by vacancy mobility rather than gas  (such as in the He case).
		
		\item Thermodynamic calculations show the PA precipitate dispersion is a kinetically trapped, non-equilibrium feature, whilst transmutation over five full-power years irreversibly redirects the alloy's chemistry (as predicted by FactSage and FISPACT-II calculations): Ni produced by Cu transmutation diverts Zr into new Ni-Zr intermetallics across the ITER operating window of \SIrange{200}{350}{\degreeCelsius}.
		
		\item The CuCrZr age-hardening strategy is therefore compromised by three compounding mechanisms governed by independent physical drivers -- ballistic (precipitate destabilisation), nuclear (compositional drift), and diffusional (gas-driven cavity nucleation) -- such that no single experimental artefact can account for all three. Degradation onset is observed within the ITER-relevant dose objective ($\approx$5~dpa) and well below anticipated DEMO lifetime doses. The apparent radiation hardening reported elsewhere \cite{fabritsiev2005effect,edwards2007effect} is attributable to defect-cluster strengthening, not retention of the original precipitate microstructure.
	\end{enumerate}
	
	Given the accelerated nature of ion irradiation, the degradation reported here should be read as a ``time-compressed'' view of the microstructural evolution expected over prolonged service rather than an artefact of the method, although quantitative transfer of degradation rates to neutron irradiation conditions will require dose-rate-aware modelling. Whilst doses of around 5 dpa within the \SIrange{150}{350}{\degreeCelsius} operational window can be considered achievable for ITER-grade CuCrZr, our results suggest that the long-term suitability of age-hardenable CuCrZr alloys for heat-sink applications may warrant renewed consideration at DEMO-relevant doses. Future work could prioritise Cu-based alloys combining high strength, thermal conductivity and irradiation resistance: recent research addresses the strength--ductility--conductivity trade-off \cite{zhang2025overcoming}, but irradiation performance remains untested in new alloys. An age-hardenable system outside Cu--Cr--Zr with low transmutation yield and irradiation-resistant precipitates should not be ruled out. A future qualification programme would require fusion-spectrum transmutation conditions, doses beyond $\approx$5 dpa, and combined H- and He-displacement exposure. In this sense, the irradiation performance of CuCrZr reported here reveals new challenges for thermonuclear fusion technology and for the continued viability of Cu-based heat-sink alloys in next-generation reactors.
	
	\section*{Acknowledgements}
	\label{acknownledgements}
	\noindent The authors gratefully acknowledge the Austrian Research Promotion Agency (FFG) for funding the Thermo Fisher Scientific Talos F200X G2 S/TEM through the project 3DnanoAnalytics (FFG-No 858040). The authors wish to thank the Engineering and Physical Sciences Research Council (EPSRC) for support for the construction of the MIAMI-2 facility (EP/M028283/1) and funding access through the UK National Ion Beam Centre (EP/X015491/1). EJM thanks the UKAEA funding of his Joint Chair position in Materials for Fusion in Birmingham through the EPSRC Energy Programme [grant number EP/ W006839/1]. CGS acknowledges the financial assistance of the Brazilian National Council of Research, Development, and Innovation (CNPq) through grants 307627/2021-7 and 305989/2024-3.  TB and MAT thank funding from the European School of Materials and all the support of Uni.-Prof. Dr. Raul Bermejo leader of the Chair of Functional Ceramics at the Montanuniversität Leoben. MAT and JT wish to express their sincere gratitude to Dr. Oliver Renk of the Chair of Physical Metallurgy, Montanuniversität Leoben, whose altruism, leadership and mentorship in materials science have shaped both their scientific development. For the purpose of open access, the authors have applied a Creative Commons Attribution (CC BY) licence to any Author Accepted Manuscript version arising.

	%\section*{References}
	\bibliographystyle{elsarticle-num}%%elsarticle-num.bst
	\bibliography{references_doi.bib}
	
\end{document}